\documentclass[intlimits,twoside,a4paper]{article}
\usepackage{graphicx}
\usepackage{wrapfig}
\usepackage{xcolor}
\usepackage[T2A]{fontenc}
\usepackage{enumerate}
\usepackage[cp1251]{inputenc}

\usepackage{cmpj3}

\usepackage{bm}

\issue{2025}{28}{1}{13801}
\doinumber{10.5488/CMP.28.13801}

\title[Entropy of restricted primitive model electrolytes]
{Entropy of a restricted primitive model electrolyte using a mean electrostatic potential approach
}

\author[{S. Lamperski}, L. B. Bhuiyan, C. W. Outhwaite,  R. Gorniak]{
\framebox{S. Lamperski}\orcid{0000-0003-4240-3724}\refaddr{label1}, L. B. Bhuiyan\orcid{0000-0002-4574-173X}\refaddr{label2}, C. W. Outhwaite\orcid{0000-0002-8825-2127}\refaddr{label3},
R. Gorniak\orcid{0000-0002-7097-6610}\refaddr{label1}  }

\addresses{
\addr{label1}Faculty of Chemistry, Adam Mickiewicz University in Pozna\'n, Uniwersytetu Pozna\'nskiego 8,
61-614 Pozna\'n, Poland
\addr{label2}Laboratory of Theoretical Physics, Department of
Physics, University of Puerto Rico, 17 Avenida Universidad, STE 1701, San Juan, Puerto
Rico 00925-2537, USA
\addr{label3}School of Mathematical and Physical Sciences, University of Sheffield,
Sheffield S3 7RH, UK
}


\date{Received November 21, 2024, in final form February 11, 2025}
\begin{document}

\maketitle
\begin{abstract}
 The excess entropy of restricted primitive model electrolytes is
calculated using a potential based approach through the symmetric Poisson-Boltzmann
and the modified Poisson-Boltzmann theories. The theories are utilized in conjunction
with a statistical thermodynamics equation that is shown to be equivalent to thermodynamic
integration. Electrolyte systems having ionic valencies 1:1 and 2:2 with diameters
$3 \times 10^{-10}$~m and 4.25 $\times 10^{-10}$~m are treated over a wide range of
concentrations.  The exact radial distribution functions for the model electrolytes
obtained from Monte Carlo simulations in the canonical ensemble are compared with the
corresponding theoretical predictions. Furthermore, the radial distribution functions
from the theories and the simulations are used in the Laird-Haymet entropy expansion
equations [ J. Chem. Phys., 1994, \textbf{100}, 3775] to estimate the excess entropy of the
solutions. These equations take into account multi-particle
distribution functions, which are approximated using a ``ring'' term. In general, the modified
Poisson-Boltzmann theory gives results that are more consistent with the simulation data
than those from the symmetric Poisson-Boltzmann theory. The results show that the excess
entropy is negative with its absolute value increasing for 1:1 electrolytes
with increasing concentration.
The symmetric Poisson-Boltzmann values are slightly overestimated, while the modified
Poisson-Boltzmann values are slightly underestimated relative to the simulations. The
curves for 1:1 electrolytes including that from the Laird-Haymet equations are consistent
with each other, while only the MPB curves for 2:2 electrolytes at $4.25 \times  10^{-10}$ m are
qualitative relative to the simulations up to about~1 mol/dm$^{3}$. The 2:2 electrolyte
curves reveal  a characteristic inflection and plateau. The results obtained in the low
concentration range~(<~0.01~mol/dm$^{3}$) are consistent with the predictions of the
Debye-H\"{u}ckel limiting law.

\printkeywords
\end{abstract}

\section{Introduction}

The concept of entropy is arguably one of the most
intriguing ones in statistical thermodynamics. The pioneering work
in relating Clausius thermodynamic entropy $S$ to the probabilistic
description of the microscopic distributions of the molecules in a
system was done by Boltzmann \cite{Boltzmann} reflected in his famous equation

\begin{equation}
	S = k_{\text{B}}\ln W ,	
\end{equation}

\noindent where $W$ is the number of available microstates with $k_{\text{B}}$ being the Boltzmann's
constant. This equation is equivalent to

\begin{equation}
	S = -k_{\text{B}}\sum _{r}p_{r}\ln p_{r} ,
	\label{eq2}
\end{equation}	

\noindent where $p_{r}$  is the probability of a distribution \cite{Gibbs}. The concept has
found its way to other areas of modern science, for example, Shannon's entropy \cite{Shannon}
in information theory has the same form as equation~(\ref{eq2}) with analogous interpretation.
Our interest in this work is more modest and we will be focusing on the entropy of
charged fluids. Entropy is one of the
more important thermodynamic functions in such systems, which is perhaps as
significant as the osmotic and activity coefficients are in characterizing the
equilibrium propeties. A standard, traditional method of
calculating entropy $S$ is through thermodynamic integration (TI) \cite{Hummer,McQuarrie}.
We will show (in the next section) that the TI
is equivalent to a known equation in statistical thermodynamics relating
the excess free energy $F^{(\text{ex})}$, the osmotic coefficient $\phi $, and the mean
activity coefficients $\gamma _{\pm}$. The use of the defining thermodynamic relation
for the free energy then yields the entropy.

    Our second approach to estimating $S$ is through the radial distribution
functions of the electrolytes. The Boltzmann equation can also be expressed
in terms of a partition function, and hence can be related to the
distribution of the constituent species. In the case of liquids, an early work
in this respect is due to Green \cite{Green} whose proposed equations turned out
to be useful. These equations involve second order (radial) and higher order
distribution functions. Subsequently, Laird and Haymet used them to calculate
the excess entropy of hard spheres \cite{Laird-Haymet1}. The contribution of higher
order distribution functions was approximated using a partial summation,
the so-called `ring' method developed by Hernando \cite{Hernando1,Hernando2}.
Very good agreement for entropy was obtained relative to that from an exact
expression emanating from the Percus-Yevick (PY) approximation for hard spheres
\cite{McQuarrie}. For electrolytes also, as in the
case of hard spheres, Laird and Haymet \cite{Laird-Haymet2} approximated the
entropy expansion up to the `ring' term. Although the equations developed
by Laird and Haymet have some limitations, they represent an important
step towards using distribution functions to calculate the entropy of liquids
and electrolytes. Silverstein, Dill and Haymet investigated the solvation entropy
using multi-molecular entropy expansion \cite{Silverstein}. Lazaridis dealt with
solvation entropy, but in heterogeneous liquids and low concentrations \cite{Lazaridis}.
It is also worth noting the work of Lazaridis and Karplus \cite{Lazaridis-Karplus}, who,
in addition to the radial distribution function, used an indicative distribution function,
which was a function of 5 angles, to describe the entropy of pure water. Recently,
Hernando and Blum proposed a new entropy component, which is related to density
fluctuations \cite{Hernando-Blum}. At the other end of the concentration scale,
we have very dilute electrolytes. Their chemical potential is described by the
well-known Debye-H\"{u}ckel limiting law (DHLL). Based on this law, Laird and Haymet
derived formulas for the limiting excess entropy and its components  \cite{Laird-Haymet2}.
The hypernetted chain (HNC) \cite{Hummer,Brian, Brian1} approximation has also been used to
calculate the electrolyte entropy.

In this study we will employ the symmetric Poisson-Boltzmann (SPB) and the
modified Poisson-Boltzmann (MPB) theories along with Monte Carlo (MC) simulations
to evaluate the $\phi$, $\gamma _{\pm}$, the energy~$U$, and hence the $S$.
We will also use the mean spherical approximation (MSA) for comparison purposes.
With regard to the alternative approach, we will utilize the radial distribution
profiles from the SPB, MPB, and the simulations together with the LH
expressions~\cite{Laird-Haymet2} to calculate $S$. In the rest of
the paper we will refer to the former procedure as TI and the latter as LH approximation.
Symmetric valency 1:1 and 2:2 electrolytes
will be treated for two different ionic sizes and for a range of concentrations.
The SPB and MPB are potential based approaches and had earlier proved to be valuable
in characterizing structural and thermodynamic properties of electrolytes and
electrolyte mixtures including colloidal solutions
\cite{Outh1,Quin1,Quin2,Molero1,Molero2}. The MSA, on the other hand, is a well
known integral equation theory with the advantage of being analytically tractable
\cite{McQuarrie}. These theories have, however,
not been previously applied to a calculation of entropy of electrolytes.
It is of interest to examine the viability of these approaches
in estimating the entropy in such systems. The SPB and MPB results will afford
useful comparative assessment of the two approaches vis-\`{a}-vis the simulations.

\section{Model and methods}
\subsection{The model}

	The physical model of the electrolyte used in this work is the restricted
primitive model (RPM). The ions are depicted as equi-sized charged hard spheres
in a structureless solvent approximated by a continuum dielectric characterized
by a dielectric constant $\varepsilon _{r}$. In the Hamiltonian, the pairwise
additive interaction potential between two ions $i$ and $j$ separated
by a distance $r$ is given by
\begin{equation}
u_{ij}(r) = \left\{ \begin{array}{cc}
\infty,
~~~~~~~~~~~~~~~~~~~~~~~~~~~~~~~ r  < d, \\
e^{2}Z_{i}Z_{j}/(4\piup \varepsilon _{0}\varepsilon _{r} r),~~~~~~ r > d ,
\end{array}\right.
\label{eq3}
\end{equation}
 where $Z_{i}$ is the valency of ion $i$, $\varepsilon _{0}$
is the vacuum permittivity, $d$ is the common ionic diameter, and $e$~is the absolute value of the fundamental charge. We have further used
symmetric 1:1 and 2:2 valencies and $d = 3 \times 10^{-10}$ m or
4.25 $\times 10^{-10}$ m. The temperature $T$ and the relative permittivity
$\varepsilon _{r}$ were held constant at 298.15 K and 78.5, respectively.

\subsection{The equivalence of thermodynamic integration with an equation
in statistical thermodynamics}

    Using TI to calculate $S$ involves the calculation of the excess free energy
$F^{(\text{ex})}$ by integrating the energy $U$. In a charged fluid system in equilibrium,
$U$ is simply the configurational potential energy. For example, the electrical part
of the free energy can be written \cite{McQuarrie,Outh1}

\begin{equation}
F^{\text{el}}=\sum _{i}e_{i}\rho _{i}\int_{0}^{1}\rd\lambda\, \psi_{i}^{\prime}(\lambda),
\label{eq4}
\end{equation}
 where $\rho _{i}$ is the mean number density of  ion species $i$, and
\begin{equation}
\psi _{i}^{\prime}=\lim _{2\rightarrow 1}\left[\psi _{i}(r_{12})-\frac{e_{i}}{4\piup \varepsilon _{0}\varepsilon _{r}r_{12}}\right],
\end{equation}
 is the potential at the centre of ion $i$ at ${\bf r}_{1}$
due to all other ions, $\psi _{i}(r_{12})$ is the mean electrostatic potential
at ${\bf r}_{2}$ given ion $i$ at ${\bf r}_{1}$
and $e_{i} = eZ_{i}$. The quantity $\lambda $ is the charging parameter such that the
charge of each ion of type $i$ at any stage of the charging process is $\lambda e_{i}$,
where $0 \leqslant \lambda \leqslant 1$ (Debye charging process).

    From thermodynamics we have

\begin{equation}
\frac{1}{\beta}\ln \gamma _{i}^{\text{el}}=\frac{\partial F^{\text{el}}}{\partial \rho _{i}}=e_{i}\int_{0}^{1}\rd\lambda\, \psi_{i}^{\prime}(\lambda),
\label{eq6}
\end{equation}
 where $\gamma _{i}^{\text{el}}$  is the electrical  part  of  the individual
activity  coefficient and $\beta=1/(k_{\text{B}}T)$. Taking $i, j = +,-$, we get from equations~(\ref{eq4}) and~(\ref{eq6})
\begin{equation}
\ln \gamma _{\pm}^{\text{el}}=\frac{\beta}{\rho}\sum _{i}\rho _{i}\frac{\partial F^{\text{el}}}{\partial \rho _{i}},
\label{eq7}
\end{equation}
 with the total number density of ions $\rho = \sum _{i}\rho _{i}$  and $\gamma _{\pm}^{\text{el}}$,
the electrical contribution to the mean activity coefficient defined through

\begin{equation}
\ln \gamma _{\pm}^{\text{el}}=\frac{1}{\rho}\sum _{i}\rho _{i}\ln \gamma _{i}^{\text{el}}.
\end{equation}

\noindent This is indeed a general definition of the mean activity coefficient (starting
from the individual activity coefficients).

    Consider now the thermodynamic expressions relating the excess osmotic coefficient
$\phi^{\text{ex}}(=\phi -1)$, the activity coefficient $\gamma _{i}$, and the excess free energy
$F^{\text{ex}}$

\begin{equation}
\phi^{\text{ex}} = -\frac{\beta F^{\text{ex}}}{\rho}+\frac{\beta}{\rho}\sum _{i}\rho _{i}\left(\frac{\partial F^{\text{ex}}}{\partial \rho _{i}}\right),
\label{eq9}
\end{equation}

\begin{equation}
\ln \gamma_{i}=\beta \left(\frac{\partial F^{\text{ex}}}{\partial \rho _{i}}\right).
\label{eq10}
\end{equation}
Equations~(\ref{eq9}) and (\ref{eq10}) lead directly to the statistical thermodynamics equation

\begin{equation}
\phi^{\text{ex}} = -\frac{\beta F^{\text{ex}}}{\rho} + \ln \gamma _{\pm}.
\label{eq11}
\end{equation}
 A similar equation has been used by Ruas et al.~\cite{Ruas} using the
binding MSA. Alternatively if the excess properties can be written as the sum of
the electrical and non-electrical parts, then equation~(\ref{eq11}) follows from equations
(\ref{eq7}), (\ref{eq9}) and (\ref{eq10}).

	Combining equation (\ref{eq11}) with the thermodynamic definition of free energy,
that is, $F = U -TS$, yields an expression for the excess entropy in the solution.

\begin{equation}
\frac{S^{(\text{ex})}}{\rho k_{\text{B}}}=\frac{\beta U}{\rho}-\ln \gamma _{\pm}+\phi ^{\text{ex}}.
\label{eq12}
\end{equation}

\noindent The left hand side of the above equation represents excess entropy per particle,
while the first term on the right hand side is the reduced energy per particle.
This equation is just another representation of TI. In passing, it is noted that in the
MSA for a PM, the thermodynamic quantities such as $\phi$, $\ln \gamma _{\pm}$, and $U$
have closed analytical forms \cite{Blum,Hoye,Sanchez-Castro}, which is useful for a quick
analysis of experimental data.

\subsection{The SPB and MPB theories}

    The symmetric Poisson-Boltzmann and the modified Poisson-Boltzmann theories are
both potential based statistical mechanical approaches to the theory of bulk electrolytes.
The details of the development of these theories have been chronicled elsewhere in the
literature \cite{Outh2,Outh3,Outh4,Outh5,Outh5ol,Outh7}. We will
restrict ourselves here to outlining the salient features of the theories along with
the relevant equations. As the term ``symmetric'' indicates, the SPB originates in the efforts,
initially by \cite{Outh2,Outh3} and later by Outhwaite and co-workers~\cite{Outh4,Outh5,Outh5ol,Outh7},
to symmetrize the radial distribution function $g_{ij}(r)$ of the conventional non-linear
Poisson-Boltzmann (PB) theory with respect to an interchange of indices, that is,
$g_{ij}(r) = g_{ji}(r)$ for asymmetric systems. In the SPB theory, the $g_{ij}(r)$ reads

\begin{equation}
g_{ij}(r)=g_{ij}^{0}(r)\exp\left\{-\frac{\beta e}{2}\left[Z_{i}(\psi _{j}(r)+\psi _{j}^{0}(r))+Z_{j}(\psi _{i}(r)+
\psi _{i}^{0}(r))\right]\right\}.
\end{equation}

\noindent Here $\psi _{i}(r)$ is the mean electrostatic potential about an ion of
species $i$ at a distance $r$, while $\psi _{i}^{0}(r) = \psi _{i}(r;Z_{i} = 0)$ is
the corresponding discharged potential. The discharged potentials are zero for a RPM
system so that the equation simplifies to

\begin{equation}
g_{ij}(r)=g_{ij}^{0}(r)\exp\left\{-\frac{\beta e}{2}\left[Z_{i}\psi _{j}(r)+Z_{j}\psi _{i}(r)\right]\right\}.
\label{eq14}
\end{equation}

The quantity $g_{ij}^{0}(r)= g_{ij}(r; Z_{} = Z_{t} =0)$ is the exclusion volume term
and is the radial distribution function between the two discharged ions in a sea of fully
charged ions.  The SPB theory for the RPM is completed by combining equation~(\ref{eq14}) with
Poisson's equation
\begin{equation}
\nabla ^{2}\psi _{i}(r)=-\frac{e}{\varepsilon _{0}\varepsilon_{r}}\sum _{j}Z_{j}\rho _{j}g_{ij}(r).
\label{eq15}
\end{equation}
The SPB has a mean field character although $g_{ij}^{0}(r)$ does contain short-range
hard-core effects. The main inter-ionic correlations occur in the neglected fluctuation potential which incorporates both the short-range and coulombic long-range contributions.

	The fluctuation potentials, missing in the SPB theory, are accounted for in the
MPB theory \cite{Outh5,Outh5ol,Outh7}, where the $g_{ij}(r)$ now transforms to
\begin{equation}
g_{ij}(r)=g_{ij}^{0}(r)\exp\left\{-\frac{\beta e}{2}\left[Z_{i}L_{i}(u_{j})+Z_{j}L_{j}(u_{i})\right]\right\},
\label{eq16}
\end{equation}
where
\begin{equation}
L_{i}(u_{j})=\frac{1}{2r(1+\kappa d)}\left\{u_{j}(r+d)+u_{j}(r-d)+\kappa \int _{r-d}^{r+d}\rd r\, u_{j}(r)\right\},
\label{eq17}
\end{equation}
\begin{equation}
\kappa = \left[\frac{\beta e^{2}}{\varepsilon_{0}\varepsilon_{r}}\sum_{i} Z_{i}^{2} \rho_{i}\right]^{\frac{1}{2}}.
\label{eq18}
\end{equation}
Here, $\kappa $ is the Debye-H\"{u}ckel constant and $u_{i}(r) = r\psi _{i}(r)$.
The equations~(\ref{eq15})--(\ref{eq18}) are the MPB equations for a RPM electrolyte.

    In both the SPB and MPB theories, the $g_{ij}^{0}(r)$ were taken to be given by
the Percus-Yevick (PY) hard sphere radial distribution functions with Verlet-Weis~(VW)
corrections \cite{Hansen-McDonald}. This has proved successful
in many previous applications of these theories to different electrolytes
\cite{Hribar,Rescic}.

\subsection{Monte Carlo simulations}

    MC simulations were performed to obtain the internal energy $U/(\rho k_{\text{B}}T)$
and the osmotic coefficient~$\phi $ in the canonical ensemble, while the
mean activity coefficient $\gamma _{\pm}$ was calculated in the inverse grand
canonical (IGCMC) ensemble \cite{Lamperski}, with the standard Metropolis algorithm being
utilized in both cases. Periodic boundary conditions and the
minimum image convention were applied in all three directions.
In the canonical ensemble, the number of ions $N$ is constant, while for IGCMC
it fluctuates. In both cases equilibration runs were
set between $(1-30)\times 10^{7}$ steps while the production runs consisted of
$(1-10)\times  10^{8}$ configurations. The interactions between ions were described
by equation~(\ref{eq3}), which includes both hard-sphere and electrostatic interactions.

 The radial distribution function $g_{ij}(r)$ can be generated using molecular
computer simulations. The results obtained are exact for the considered model and
can serve as a reference point for assessing an approximate theory. In statistical
thermodynamics, $g_{ij}(r)$ is defined as the probability of finding a molecule~$j$
located at a distance $r$ from a molecule $i$, called the central one. This definition
is mathematically described by the equation

\begin{equation}
g_{ij}(r)=\frac{\rd N_{j}(r)}{\rd V(r)\rho _{j}}	,
\label{eq19}
\end{equation}
 where d$N_{j}$ is the average number of molecules $j$ in a spherical shell with radius $r$,
thickness d$r$ and volume d$V$. In computer simulations, $g_{ij}(r)$ is calculated from the
above formula. A typical algorithm can be found in the Frenkel and Smit textbook \cite{Frenkel}.
The calculations were carried out for a wide range of electrolyte concentrations
from 0.001 mol/dm$^{3}$ to  6.76 mol/dm$^{3}$. The canonical ensemble simulations
utilized 5000 ions at low concentrations with a step-length (in units scaled with respect
to the side length of the simulation box) close to unity. However, at high concentrations,
these values were 2000 and 0.05, respectively. Large numbers of ions ensured that the $g_{ij}(r)$
functions converged to 1 at large distances, which is a necessary condition to obtaining
correct entropy data.

\subsection{Laird-Haymet entropy expansions}

The pair distributions $g_{ij}(r)$'s obtained from the MC, SPB, and MPB theories were
used in the entropy expansion, in terms of multi-particle correlations, developed
initially by Hernando \cite{Hernando1,Hernando2} and later by Laird and Haymet
\cite{Laird-Haymet2} to calculate the entropy. The LH formalism entails
calculation of the quantities $S^{(2)}$, $S_{\text{ring}}$, both of which involve pair
correlations, with the excess entropy being written as

\begin{equation}
S^{(\text{ex})} = S^{(2)} + S_{\text{ring}},
\label{eq20}
\end{equation}

\noindent the two terms on the right-hand side being the second and third terms in the entropy expansion.
In particular, $S_{\text{ring}}$ may be taken to be an indirect contribution of higher order terms.

    {For completeness, we quote here the specific expressions for $S^{(2)}$ and $S_{\text{ring}}$
\cite{Laird-Haymet2} using the present notations:

\begin{equation}
S^{(2)} = -\frac{1}{2\rho}\sum _{i}\sum _{j}\rho _{i}\rho _{j}\int \rd{\bf {r}}_{ij}\left[g_{ij}(r_{ij})\ln g_{ij}(r_{ij})-g_{ij}(r_{ij})+1\right],
\label{eq21}
\end{equation}
and
\begin{equation}
S_{\text{ring}}=\frac{1}{2\rho (2\piup )^{3}}\int \rd\textbf{k}\left[\ln |\textbf{I}+\tilde{\textbf{H}}(k)|+\frac{1}{2}\textbf{Tr} \left(\tilde{\textbf{H}}^{2}(k)\right)-\textbf{Tr}\left(\tilde{\textbf{H}}(k)\right)\right].
\label{eq22}
\end{equation}
Here, \textbf{H} and \textbf{I} are 2$\times$2 matrices, with the latter being the identity matrix.
\begin{equation}
\tilde{\textbf{H}}(k)_{ij} = \rho _{i}^{1/2}\rho _{j}^{1/2}\tilde{h}_{ij}(k),
\label{eq23}
\end{equation}
 where $\tilde{h}_{ij}(k)$ is the Fourier transform of $h_{ij}(r)=g_{ij}(r)-1$,
and for an isotropic, homogeneous fluid, $\tilde{h}_{ij}(k)$ can be written as

\begin{equation}
\tilde{h}_{ij}(k) = 4\piup \int _{0}^{\infty}\rd r r^{2}\left\{g_{ij}(r)-1\right\}\frac{\sin kr}{kr}.
\label{eq24}
\end{equation}
\begin{figure}[!t]
	\begin{center}
		\includegraphics[width=0.55\textwidth]{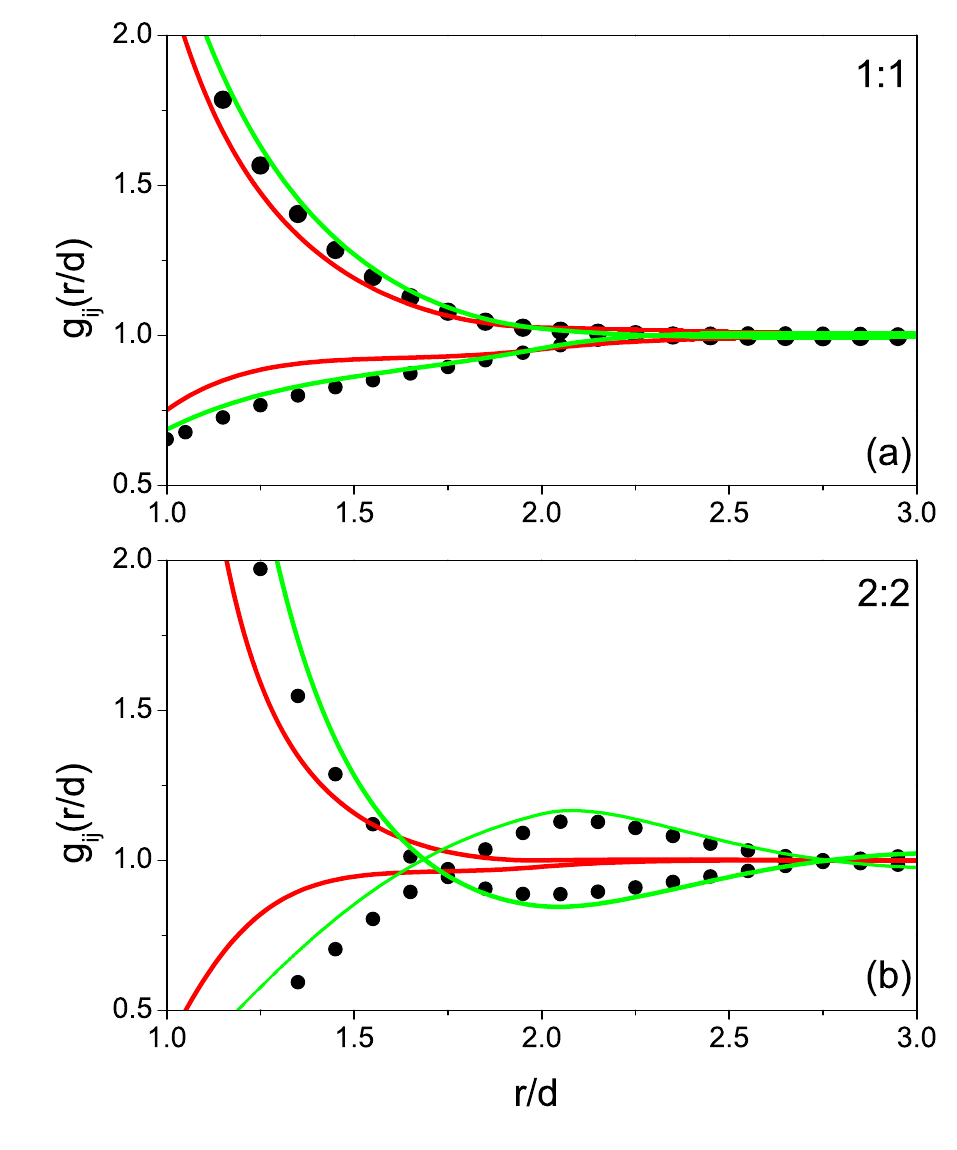}
		\caption{(Colour online) Radial distribution function, $g_{ij}(r/d)$ as a
			function of $r/d$  for  a  RPM  electrolyte, $d =$ 4.25 $\times $ 10$^{-10}$ m, $c =$ 1.96
			mol/dm$^{3}$,
			$T =$ 298.15 K, and  $\varepsilon _{r} =$ 78.5.
			Legend:   SPB (red lines), MPB (green lines), and MC (black filled circles).
			Panel (a) 1:1 valency, panel (b) 2:2 valency.}
		\label{fig1}
	\end{center}
\end{figure}

    For numerical integration purposes, the right-hand sides of equations~(\ref{eq21}) and (\ref{eq22}) were
expanded and cast in the following forms. For example, after expanding the right-hand side
of equation~(\ref{eq21}), it can be written as
\begin{equation}
S^{(2)} = I_{1}+I_{2}+I_{3},
\label{eq25}
\end{equation}
\begin{equation}
I_{1}=-3\eta \int_{0}^{\infty}\rd r_{ii}^{*}(r_{ii}^{*})^{2}\left\{g_{ii}(r^{*})\ln g_{ii}(r^{*})-g_{ii}(r^{*})+1\right\},
\label{eq26}
\end{equation}
\begin{equation}
I_{2}=-6\eta \int_{0}^{\infty}\rd r_{ij}^{*}(r_{ij}^{*})^{2}\left\{g_{ij}(r^{*})\ln g_{ij}(r^{*})-g_{ij}(r^{*})+1\right\},
\label{eq27}
\end{equation}
 $I_{3}$ is now obtained from $I_{1}$ simply by the interchange $i \leftrightarrow j$.
In the above equations, $\eta = \frac{\piup}{6}\rho a^{3}$ is the total packing fraction, and $r^{*} = r/a$.
Defining further, $y = ka$, and $M_{ij}(k)=\tilde{h}_{ij}(k)/a^{3}$, equation (\ref{eq22}) becomes

\begin{equation}
S_{\text{ring}}=I_{4}+I_{5}+I_{6},
\label{eq28}
\end{equation}
\begin{equation}
I_{4}=\frac{1}{24\piup \eta}\int_{0}^{\infty}\rd y y^{2}\ln\left\{1+\left(\frac{6\eta}{2\piup}\right)^{2}M_{ii}M_{jj}
+\left(\frac{6\eta}{2\piup}\right)(M_{ii}+M_{jj})-\left(\frac{6\eta}{2\piup}\right)^{2}M_{ij}^{2}\right\},
\label{eq29}
\end{equation}

\begin{equation}
I_{5}=\frac{1}{24\piup \eta}\int_{0}^{\infty}\rd y y^{2}\left\{\frac{1}{2}\left(\frac{6\eta}{2\piup}\right)^{2}\left(M_{ii}^{2}+2M_{ij}^{2}+M_{jj}^{2}\right)\right\},
\label{eq30}
\end{equation}
 and
\begin{equation}
I_{6}=-\frac{1}{24\piup \eta}\int_{0}^{\infty}\rd y y^{2}\left(\frac{6\eta}{2\piup}\right)\left(M_{ii}+M_{jj}\right).
\label{eq31}
\end{equation}
\begin{figure}[!t]
	\begin{center}
		\includegraphics[width=0.55\textwidth]{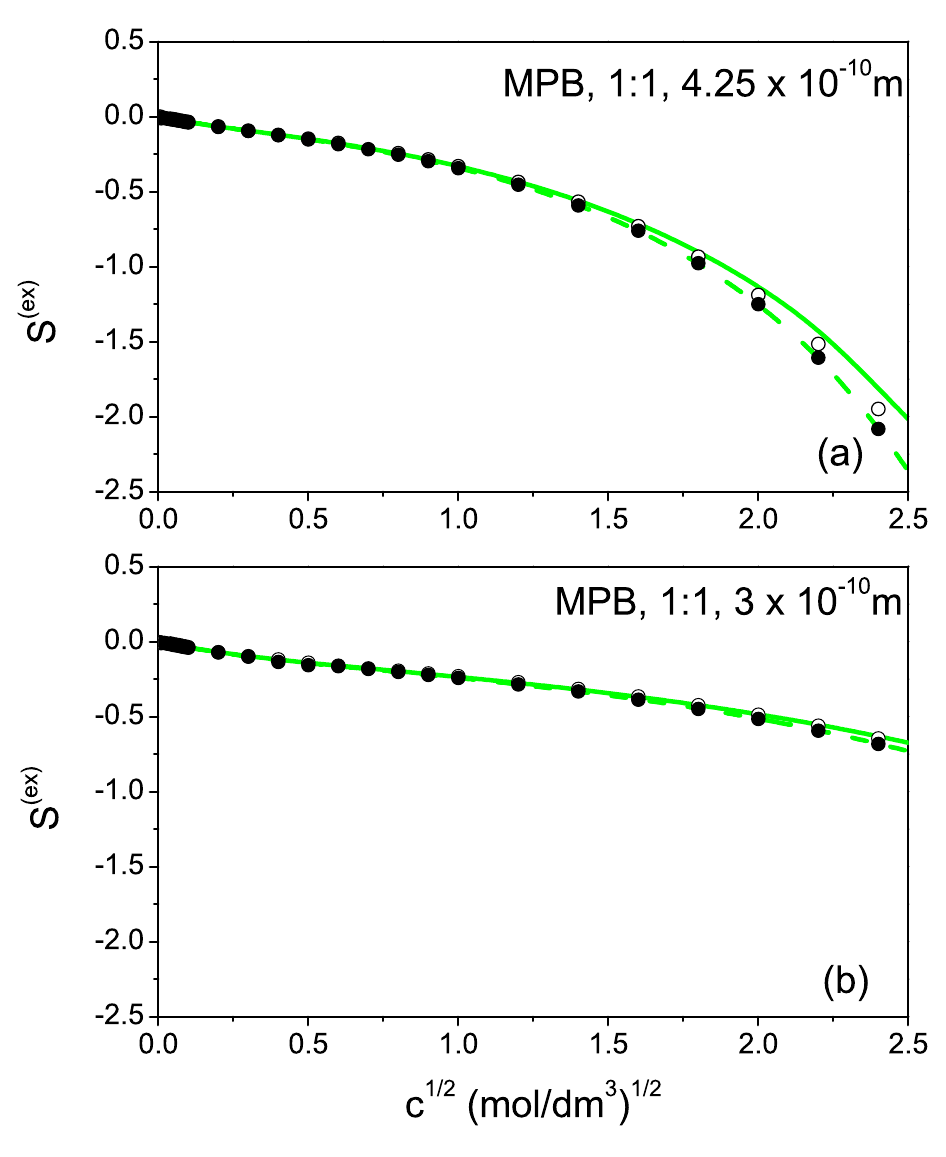}
		\caption{(Colour online) Excess entropy $S^{(\text{ex})}$ as a function of the
			square root of electrolyte concentration for a 1:1 valency RPM electrolyte.
			{Legend: MPB$_{TI}$ (solid green lines), MPB$_{LH}$ (dashed green lines), MC$_{TI}$ (black filled),
				and MC$_{LH}$ (black open circles).}
			Panel (a) $d=4.25 \times  10^{-10}$ m, Panel (b) $d = 3 \times 10^{-10}$ m.
			The SPB and MPB predictions are only distinguishable on the graphical scale at the highest
			concentrations. The remaining RPM parameters as in figure~\ref{fig1}.}
		\label{fig2}
	\end{center}
\end{figure}
\begin{figure}[!t]
	\begin{center}
		\includegraphics[width=0.55\textwidth]{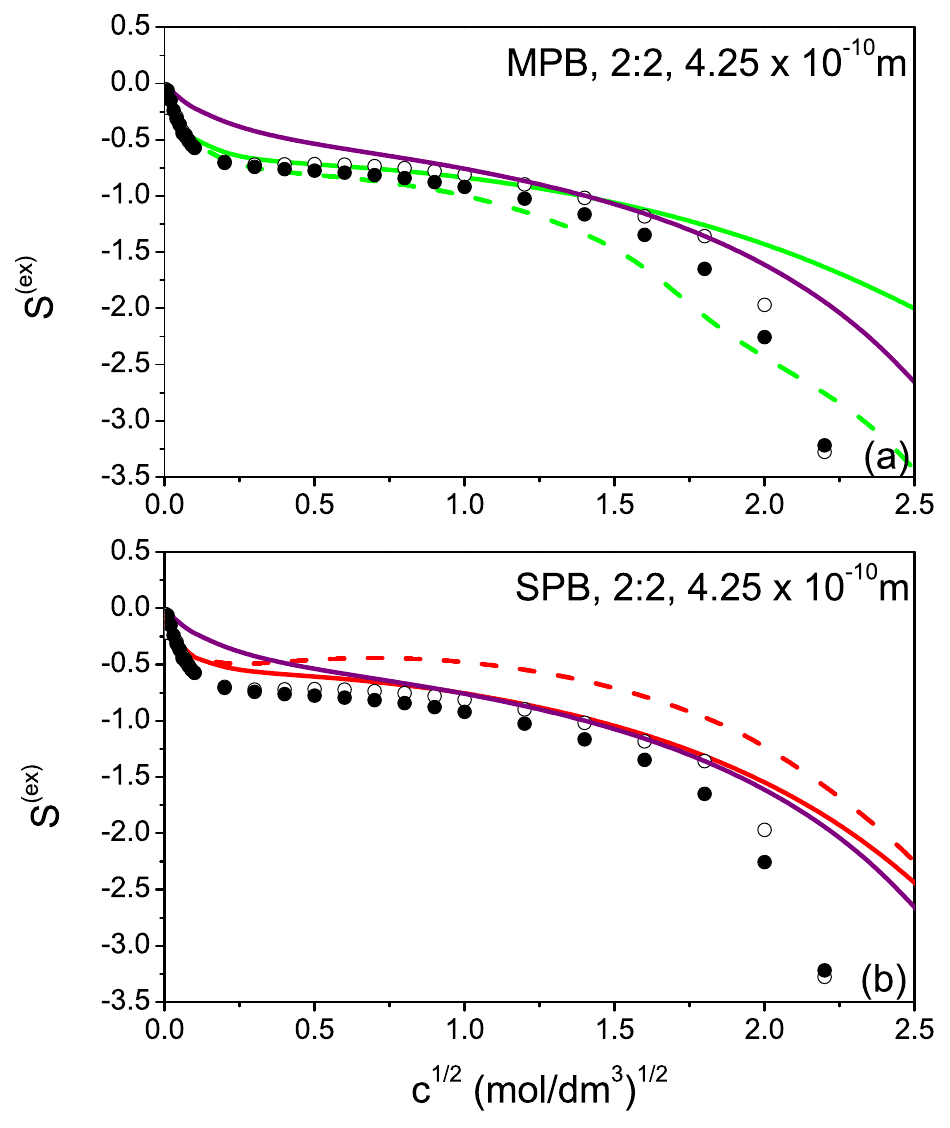}
		\caption{(Colour online) Excess entropy $S^{(\text{ex})}$ as a function of the
			square root of electrolyte  concentration  for  a 2:2  valency RPM electrolyte,
			$d = 4.25 \times 10^{-10}$ m.
			{Legend: panel (a) MPB$_{TI}$ (solid green lines),  MPB$_{LH}$ (dashed green lines), MSA$_{TI}$ (purple line),
				MC$_{TI}$ (black filled circles), and MC$_{LH}$ (black open circles). Panel (b) SPB$_{TI}$ (solid red lines),
				SPB$_{LH}$ (dashed red lines), MSA$_{TI}$ (purple line), MC$_{TI}$ (black filled circles), and MC$_{LH}$ (black
				open circles)}. The remaining RPM parameters as in figure~\ref{fig1}.}
		\label{fig3}
	\end{center}
\end{figure}
\begin{figure}[!t]
	\begin{center}
		\includegraphics[width=0.5\textwidth]{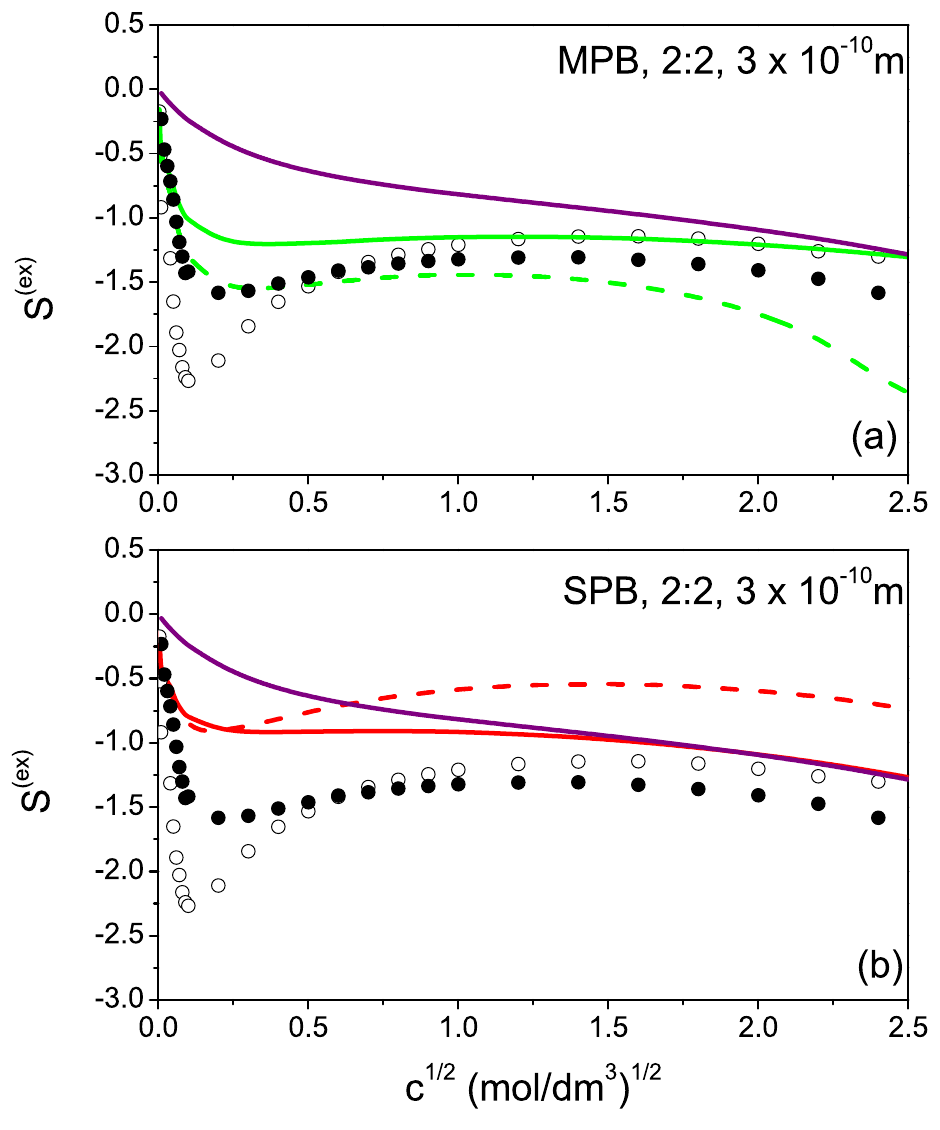}
		\caption{(Colour online) Excess entropy $S^{(\text{ex})}$ as a function of the
			square root of electrolyte concentration in a 2:2 valency RPM electrolyte, $d = 3 \times 10^{-10}$ m.
			{Legend: panel (a) MPB$_{TI}$ (solid green lines),  MPB$_{LH}$ (dashed green lines), MSA$_{TI}$ (purple line),
				MC$_{TI}$ (black filled circles), and MC$_{LH}$ (black open circles). Panel (b) SPB$_{TI}$ (solid red lines),
				SPB$_{LH}$ (dashed red lines), MSA$_{TI}$ (purple line), MC$_{TI}$ (black filled circles), and MC$_{LH}$ (black
				open circles).} The remaining  RPM  parameters  as  in figure~\ref{fig1}.}
		\label{fig4}
	\end{center}
\end{figure}

The SPB, MPB, or MC results for $S^{(2)}$, $S_{\text{ring}}$, and hence $S^{(\text{ex})}$ are
now obtained by using the relevant pair correlation functions $g_{ij}$ in the expressions
for $I_{1} -I_{6}$ with the integrals being evaluated by using the Simpson rule.
As a check on the numerics we could reproduce some of the results in table~1 of reference~\cite{Laird-Haymet1} for a pure hard sphere fluid with the $g_{ij}$ given by the PY + VW theories.}

    Using the HNC equation, Laird and Haymet \cite{Laird-Haymet2} showed that truncating the
expansion beyond the `ring' term was a viable approximation up to moderate solution concentrations.
The DHLL values of $S^{(2)}$  and $S_{\text{ring}}$ are

\begin{equation}
S^{(2)} = -\frac{\kappa ^{3}}{32\piup \rho},
\label{eq32}
\end{equation}
and
\begin{equation}
S_{\text{ring}} = -\frac{\kappa ^{3}}{96 \piup \rho}
\label{eq33}
\end{equation}
with their sum giving the correct DHLL value of
\begin{equation}
S^{(\text{ex})}=-\frac{\kappa ^{3}}{24 \piup \rho}.
\label{eq34}
\end{equation}

Retention of $S_{\text{ring}}$ in equation~(\ref{eq20}) is thus required to give the requisite DHLL value
\cite{Laird-Haymet2}. These DHLL expressions are linear with respect to $\sqrt{c}$,
and at a given concentration, depend only on the ionic valency, being independent of ionic size.

\section{Results and discussion}

	In the theoretical SPB, MPB, and MSA calculations, and in the MC simulations,
we have used the following physical parameters for the RPM electrolyte. Two different
values for the common diameter of the ions were taken,
$3 \times 10^{-10}$ m and $4.25 \times 10^{-10}$ m. Previously,
we had worked with the latter value,
which to some extent takes into account the hydration of ions. However, the smaller
diameter of the ion results in an almost 3-fold reduction in the volume of the ion,
which translates into a decrease in steric interactions. To assess the effect of
interionic electrostatic interactions on the entropy, we have also used two different
valencies, 1:1 and 2:2, respectively. A 2-fold increase in the charge of an ion is
accompanied by a 4-fold increase in electrostatic interactions. Thus, we have four
all symmetric electrolyte models that differ in ionic diameter and in ionic valency.
In model A, the diameter was $d = 4.25 \times  10^{-10}$ m and the valency $Z = \pm $1.
In model B: $d = 4.25 \times  10^{-10}$ m, $Z = \pm $2, in C: $d = 3 \times  10^{-10}$ m,
$Z = \pm $ 1, and in D: $d = 3 \times  10^{-10}$ m, $Z = \pm $ 2. The theoretical
calculations and the simulations were performed for 27 electrolyte concentrations
ranging from $c = 0.0001$~mol/dm$^{3}$ to $c = 6.76$~mol/dm$^{3}$.
\begin{figure}[!t]
	\begin{center}
		\includegraphics[width=0.5\textwidth]{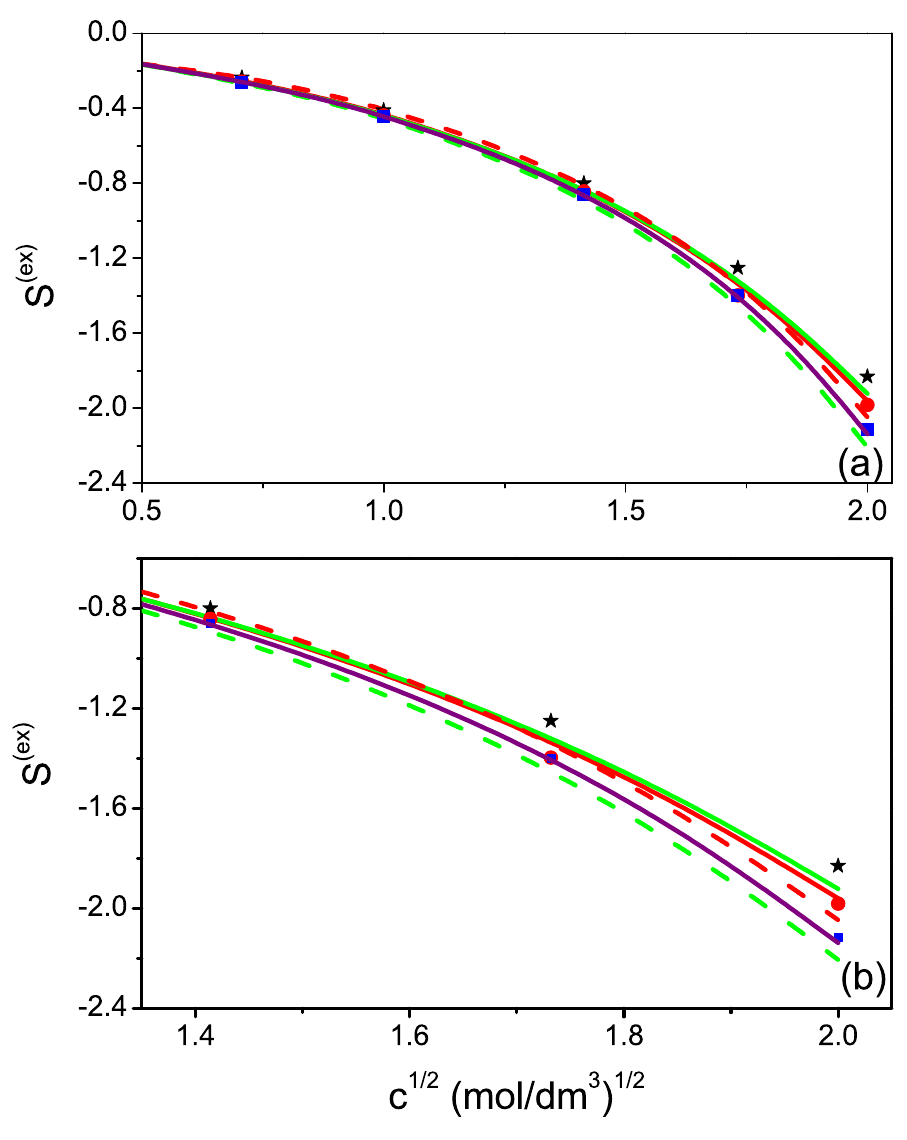}
		\caption{(Colour online) Excess entropy $S^{(\text{ex})}$  as a function of the
			square root of an  electrolyte  concentration for  a  1 : 1  RPM  electrolyte ,
			$d =$ 4.9 $\times$ 10$^{-10}$ m, $T =$ 298 K, and  $\varepsilon _{r} =$ 78.356.
			{Legend: SPB$_{TI}$ (solid red line), MPB$_{TI}$ (solid green line), SPB$_{LH}$ (dashed red line),
				MPB$_{LH}$ (dashed green line), reference~\cite{Hummer} (4th column of table IV) (black stars),
				reference~\cite{Hummer} (MC Widom) (red filled circles)}, reference \cite{Hummer} (HNC) (blue squares).
			Panel (a) $c^{1/2}$ ranges from 0.5 (\text{mol/dm}$^{3})^{1/2}$ to 2$(\text{mol/dm}^{3})^{1/2}$, panel
			(b) $c^{1/2}$ ranges from 1.35 $(\text{mol/dm}^{3})^{1/2}$ to 2 $(\text{mol/dm}^{3})^{1/2}$.}
		\label{fig5}
	\end{center}
\end{figure}
\begin{figure}[!t]
	\begin{center}
		\includegraphics[width=0.5\textwidth]{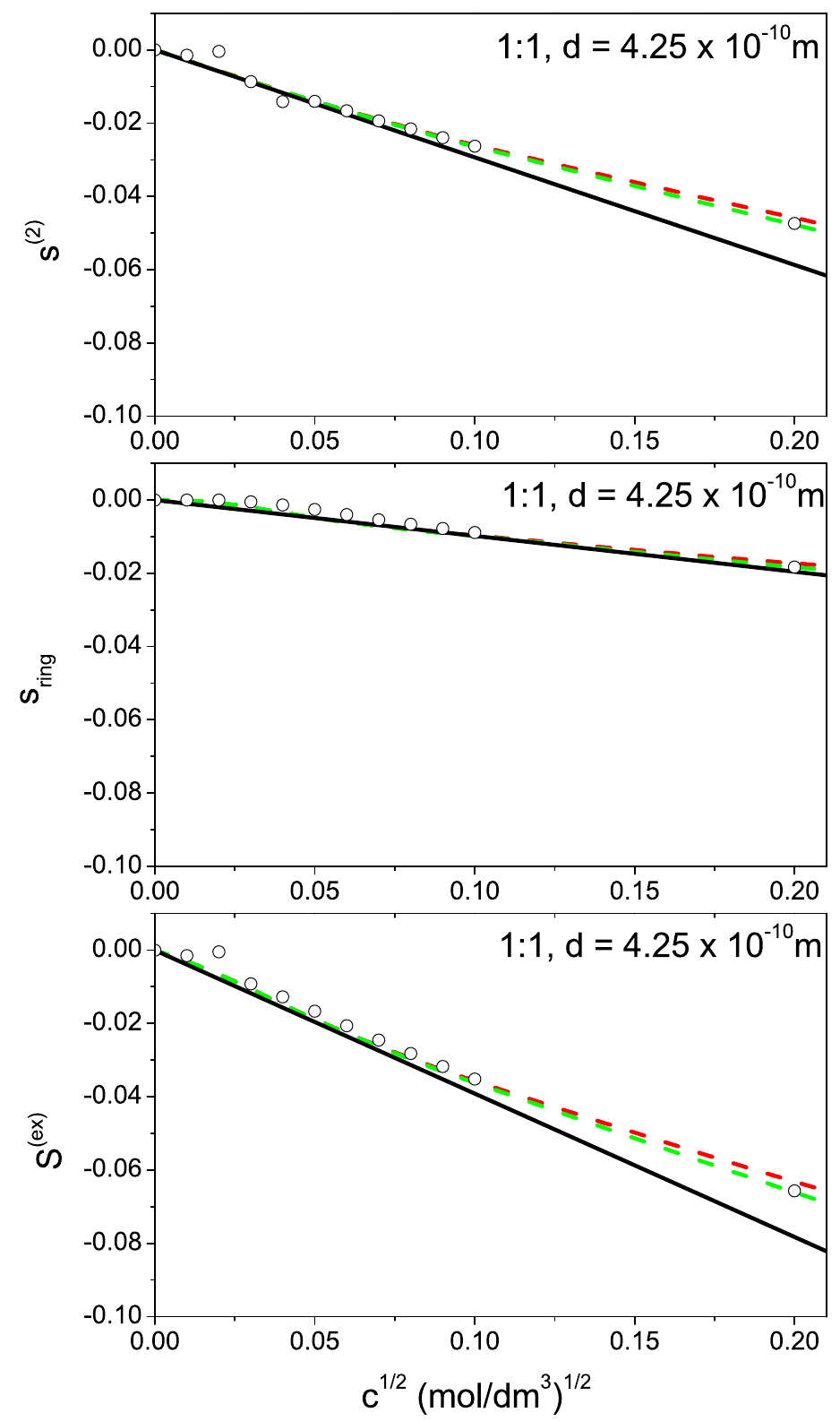}
		\caption{(Colour online) Excess entropy $S^{(\text{ex})}$, and its components,
			$S^{(2)}$, and, $S_{\text{ring}}$, as functions of the square root of electrolyte concentration
			for a 1:1 RPM electrolyte with ion diameter $d = 4.25\times 10^{-10}$~m at a low concentration regime.
			{Legend: SPB$_{LH}$ (dashed red lines), MPB$_{LH}$ (dashed green lines), and MC$_{LH}$ (black open circles).
				DHLL (solid black line).} The rest of the physical parameters as in figure~\ref{fig1}.}
		\label{fig6}
	\end{center}
\end{figure}
\begin{figure}[!t]
	\begin{center}
		\includegraphics[width=0.47\textwidth]{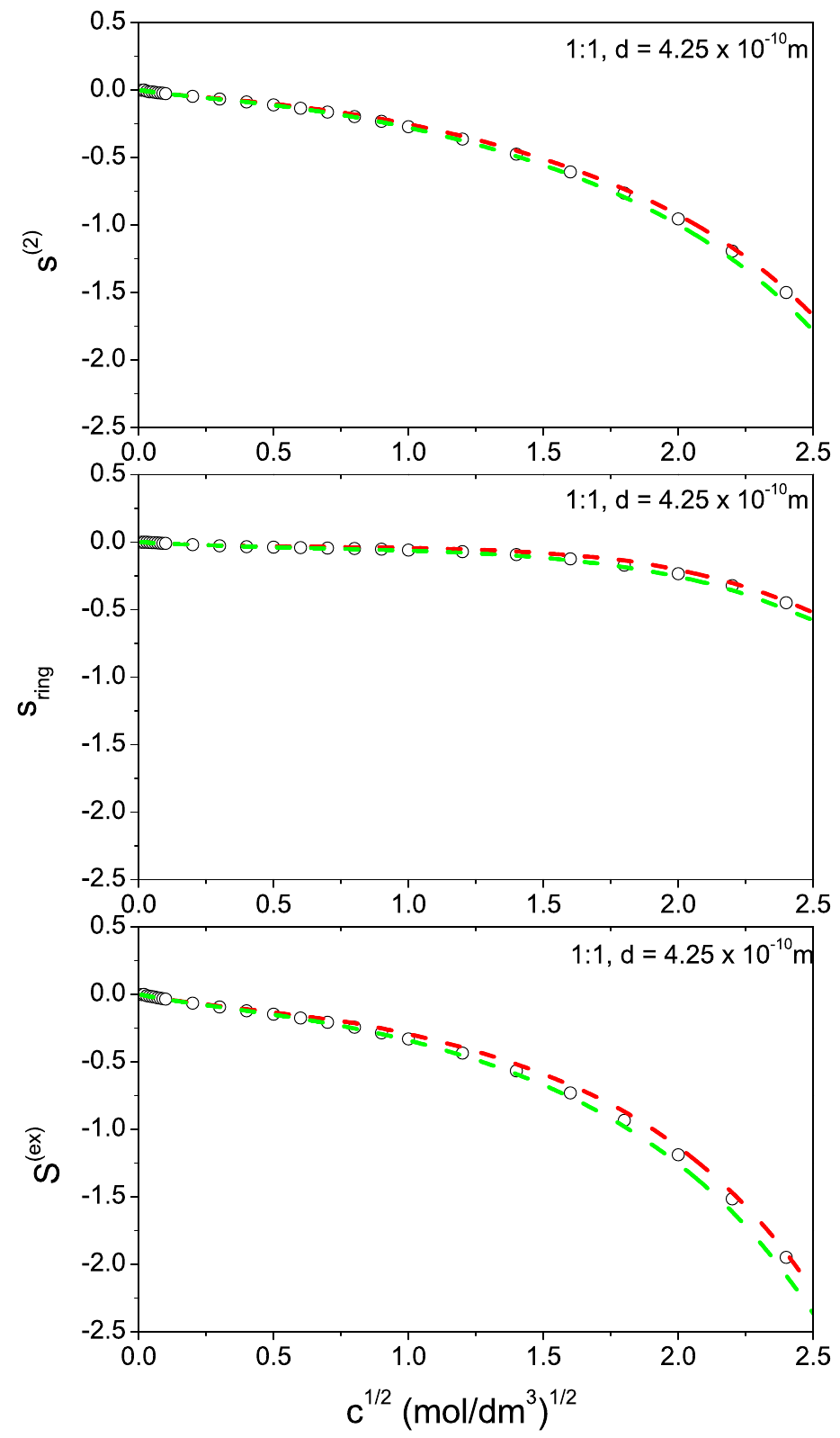}
		\caption{(Colour online) The same as figure~\ref{fig6}, but for a broader concentration regime.
			DHLL not shown here.}
		\label{fig7}
	\end{center}
\end{figure}
\begin{figure}[!t]
	\begin{center}
		\includegraphics[width=0.48\textwidth]{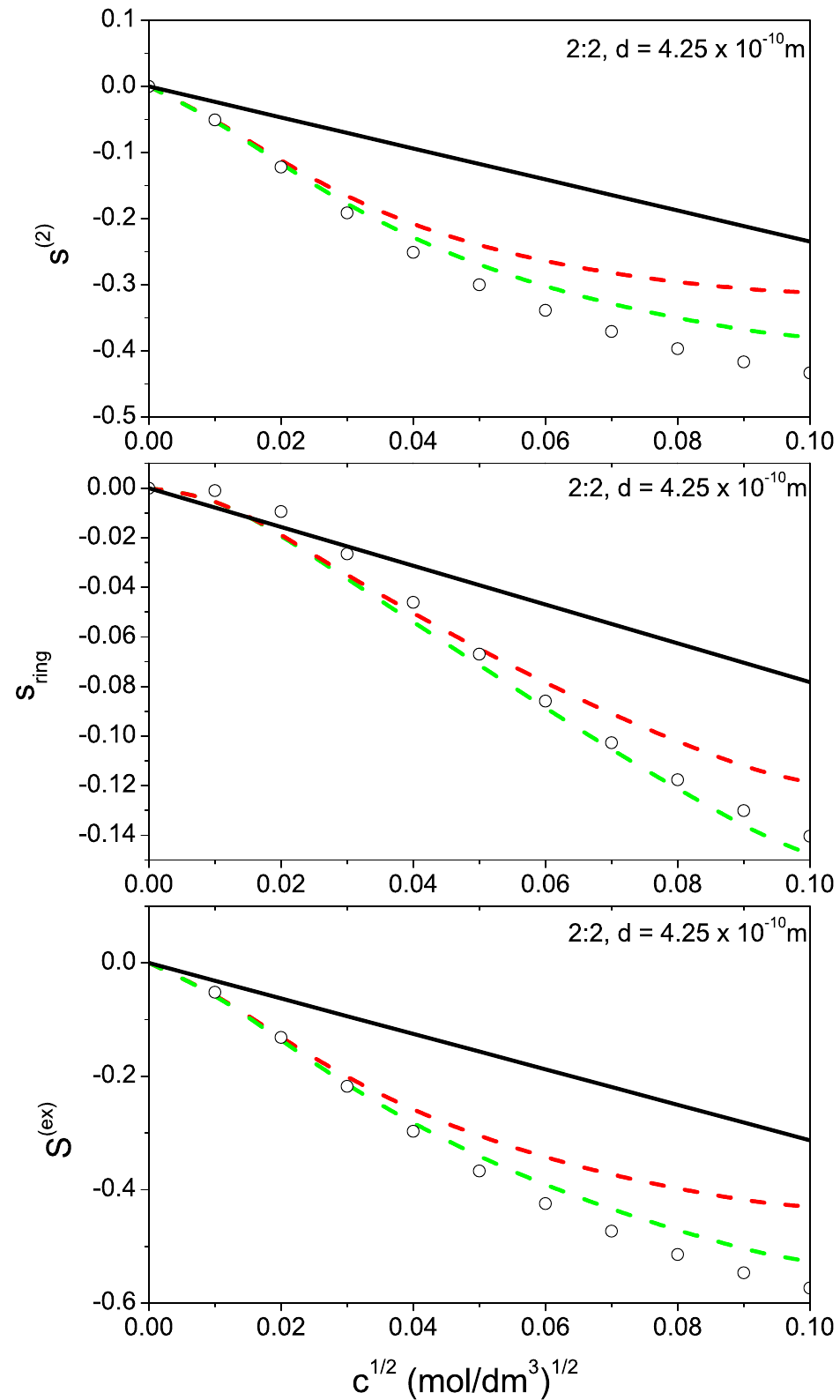}
		\caption{(Colour online) Excess entropy $S^{(\text{ex})}$ and its components,
			$S^{(2)}$ and $S_{\text{ring}}$, as functions of the square root of electrolyte concentration
			for a 2:2 RPM electrolyte, $d = 4.25 \times 10^{-10}$ m, at a low concentration regime.
			{Legend: SPB$_{LH}$ (dashed red lines), MPB$_{LH}$ (dashed green lines), and MC$_{LH}$ (black open circles).
				DHLL (solid black line).} The rest of the physical parameters as in figure~\ref{fig1}.}
		\label{fig8}
	\end{center}
\end{figure}

    A basic quantity in our research is the radial distribution function
$g_{ij}(r)$ and so we begin this discussion by illustrating some structural
results manifest through $g_{ij}(r)$ at some selected concentrations.
Figure~\ref{fig1} shows the SPB and MPB distribution functions at
$c =$ 1.96 mol/dm$^{3}$, while the symbols refer to the MC data.
The figure is for a RPM system with $d = 4.25 \times $ 10$^{-10}$~m,
and 1:1 valency (upper panel) and 2:2 valency (lower panel), respectively. The
thickness of the ionic atmosphere is over three times the ion diameter with the MC
and MPB ionic profiles $g_{ij}(r)$ displaying damped oscillations for both the systems.
Although the damped oscillations for the 1:1 case are relatively faint, they are
quite substantial for the higher 2:2 valency case. Such damped oscillations are
typical of charged fluid systems. In electric double layers, they can lead to
overcharging of the electrode.  At this high concentration, the MPB profiles are
still closely following their MC counterparts. However, the mean field SPB theory
does not capture this effect with the SPB $g_{ij}(r)$  being monotonously decreasing
and are thus not qualitative. The behaviour of the $g_{ij}(r)$ for the other two
models involving $d = 3 \times $ 10$^{-10}$~m are not shown as they show similar
trends to  $d =4.25 \times $ 10$^{-10}$~m.

    In figures \ref{fig2}--\ref{fig4} we show the TI results for the excess entropy $S^{(\text{ex})}$
as function of $\sqrt{c}$ from the SPB, MPB, MSA, and MC along with the corresponding
results from the LH approximations. To avoid confusion, we  use
the notation X$_{TI}$ and X$_{LH}$ to denote the two cases where X stands for SPB, MPB, MSA, or MC.
Figure~\ref{fig2} illustrates the MPB$_{TI}$ (green curves), MPB$_{LH}$ (green dashed curves),
MC$_{TI}$ (black filled circles), and MC$_{LH}$ (black open circles) for a 1:1 electrolyte at
$d = 4.25 \times 10^{-10}$ m (upper panel) and $d = 3 \times  10^{-10}$ m (lower panel).
As can be seen, in either case, the curves and the symbols are remarkably consistent with
each other with hardly any visible difference among them. The corresponding SPB and MSA
curves lie on top of the MPB curves and are hence not displayed. We thus have that for
the 1:1 valency (models A and C), the theories are capable of predicting the simulation data
almost quantitatively up to a fairly high concentration.
\begin{figure}[!t]
	\begin{center}
		\includegraphics[width=0.5\textwidth]{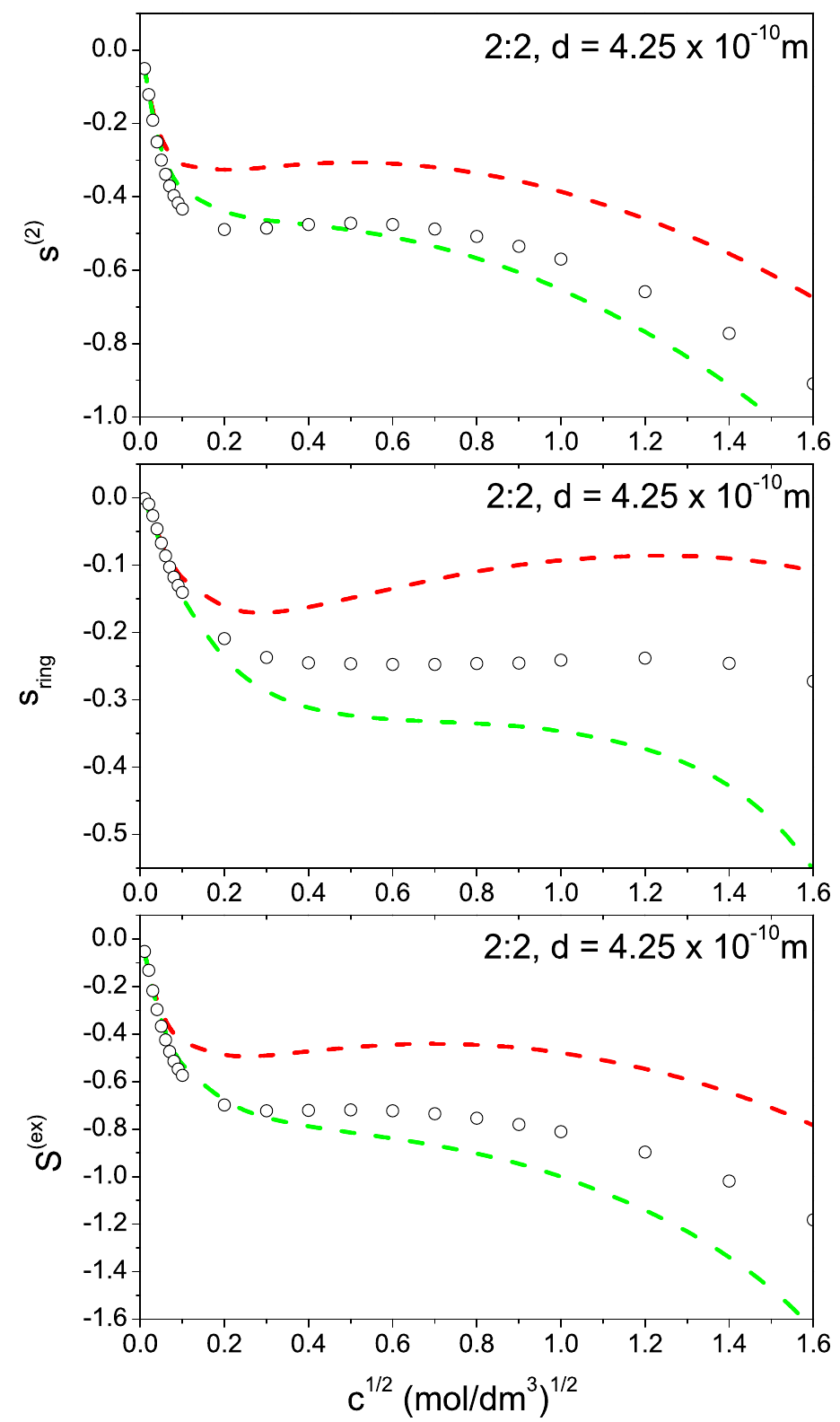}
		\caption{(Colour online) The same as figure~\ref{fig8}, but for a broader concentration
			regime. DHLL not shown here.}
		\label{fig9}
	\end{center}
\end{figure}
\begin{figure}[!t]
	\begin{center}
		\includegraphics[width=0.5\textwidth]{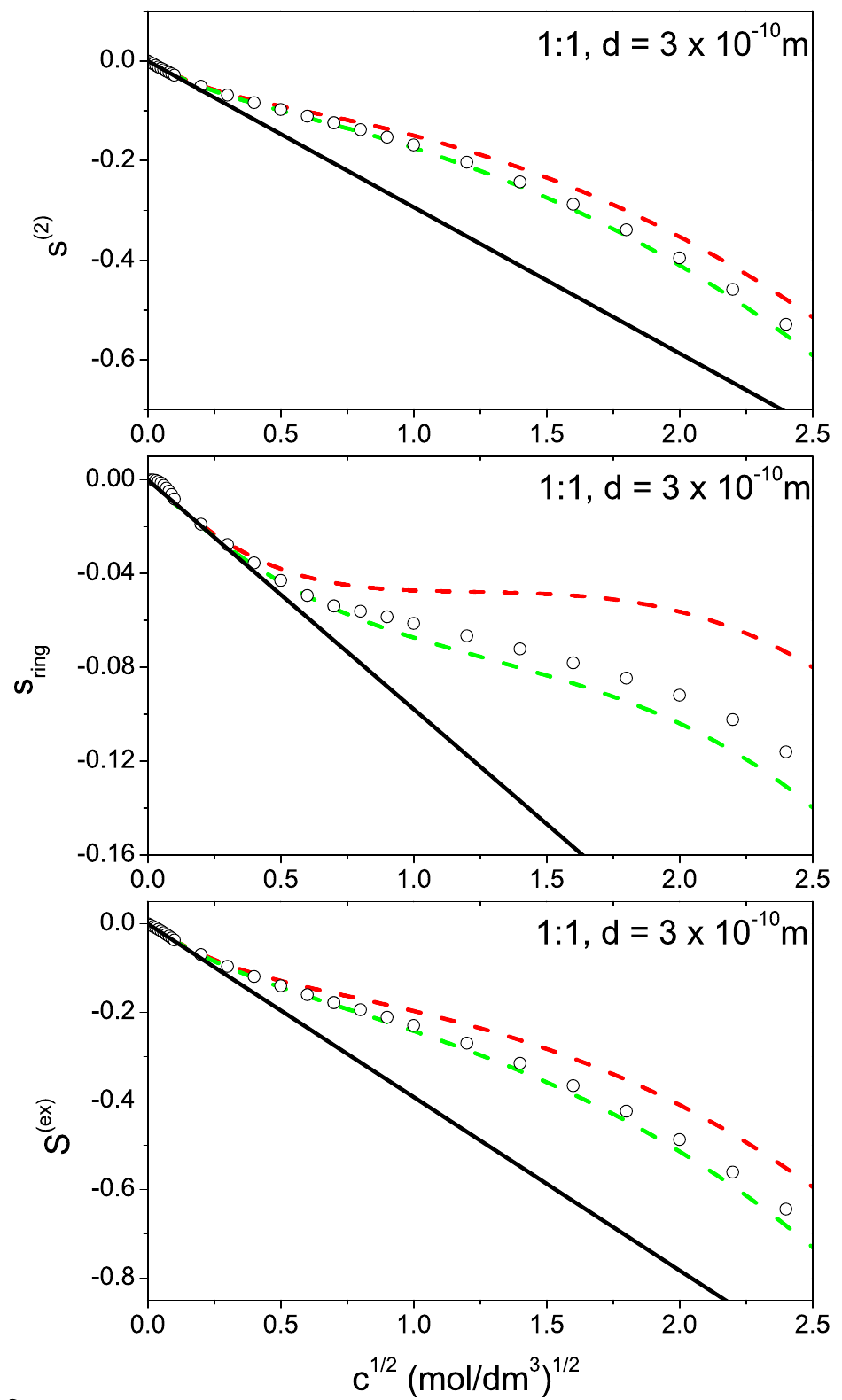}
		\caption{(Colour online) Excess entropy $S^{(\text{ex})}$ and its components
			$S^{(2)}$ and $S_{\text{ring}}$, as functions of the square root of electrolyte concentration
			for a 1:1 RPM electrolyte, $d = 3 \times 10^{-10}$~m.
			{Legend: SPB$_{LH}$ (dashed red lines), MPB$_{LH}$ (dashed green lines), and
				MC$_{LH}$ (black open circles). DHLL (solid black line).} The rest of the physical parameters
			as in figure~\ref{fig1}.}
		\label{fig10}
	\end{center}
\end{figure}

    The situation changes substantially in figures~\ref{fig3} and \ref{fig4}, which show the $S^{(\text{ex})}$
for the 2:2 valencies at $d = 4.25 \times 10^{-10}$ m and $d = 3 \times 10^{-10}$ m,
respectively. Although in figure~\ref{fig3} the theories are still broadly qualitative, the
SPB$_{LH}$ and MPB$_{LH}$ now reveal discrepancies from the simulation data, especially
at higher concentrations. With the exception of the MSA$_{TI}$, the theories and the
MC$_{TI}$, MC$_{LH}$ reveal a plateau around $c \sim  0.06$ mol/dm$^{3}$. The MSA$_{TI}$
curve, on the other hand, decreases monotonously. Some differences between the two sets
of MC data also emerge at higher concentrations. This is possibly due to the approximations
in the LH entropy expansion scheme since the same $g_{ij}(r)$'s are used in the
two cases. The picture changes again in figure~\ref{fig4} at $d = 3 \times 10^{-10}$ m, where the
MC$_{LH}$ shows a deep minimum at low concentration (around $c \sim  0.002$~mol/dm$^{3}$,
while the MC$_{TI}$, SPB$_{TI}$, SPB$_{LH}$, MPB$_{TI}$, MPB$_{LH}$ have a shallow minimum.
The MSA$_{TI}$ continues to be monotonously decreasing as in figure~\ref{fig3}. It is likely that a
simultaneous increase in Coulombic interactions and a decrease in steric interactions lead
to a less random distribution of particles at very low concentrations and hence to a decrease
in entropy. The MPB does not give the $d = 0$ (zero ion size) limiting behaviour at low
concentrations \cite{Kjellander,Mitchell}, while for the planar electric double layer it
was suggested that the MPB decreases in accuracy as the ion diameter is reduced
\cite{Carnie}. Although perhaps not being the reason for the apparent ``poor'' behaviour for 2:2,
it is an indication that the MPB loses accuracy as ion size decreases at low concentrations.
\begin{figure}[!t]
	\begin{center}
		\includegraphics[width=0.5\textwidth]{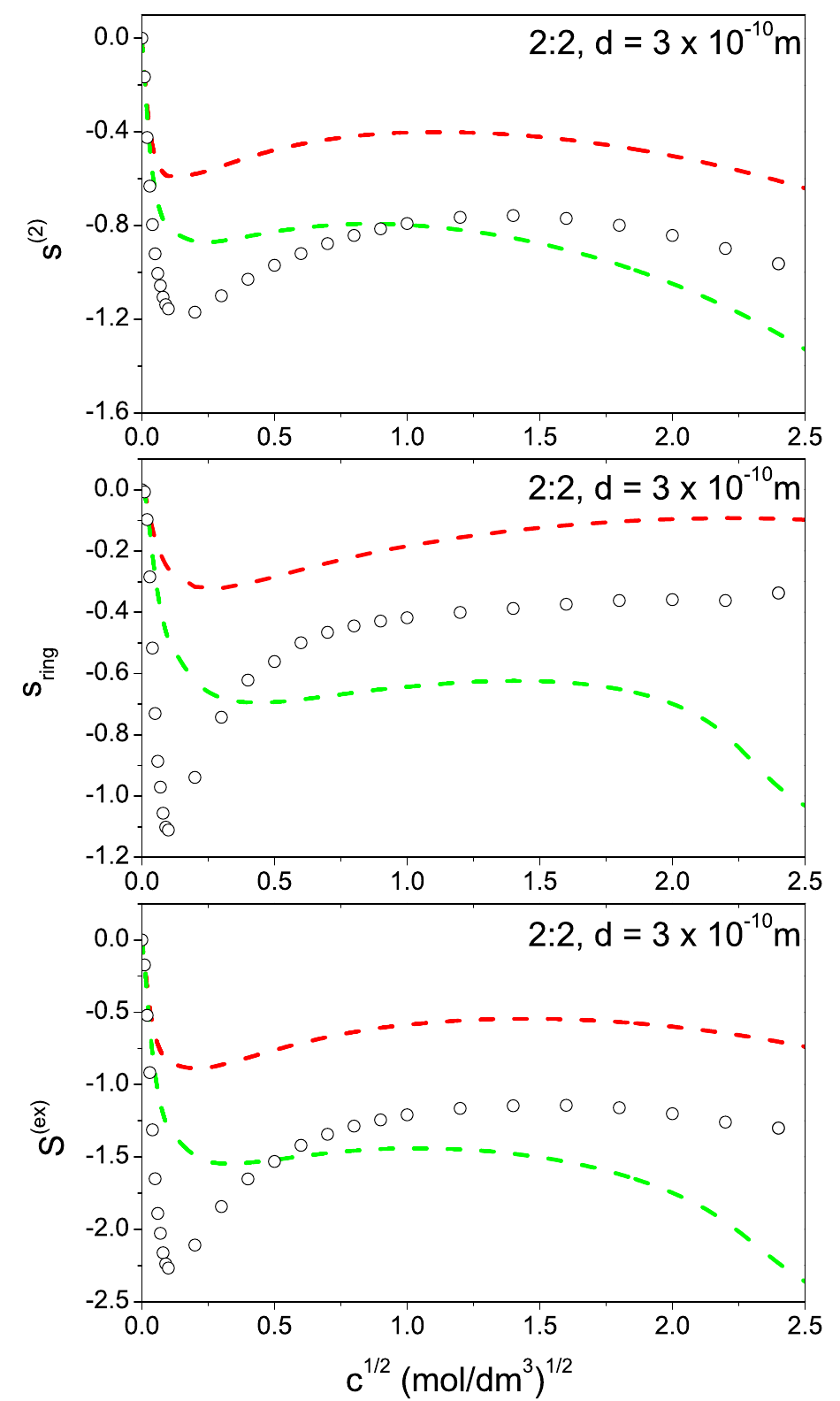}
		\caption{(Colour online) Excess entropy $S^{(\text{ex})}$ and its components
			$S^{(2)}$ and $S_{\text{ring}}$, as functions of the square root of electrolyte concentration
			for a 2:2 RPM electrolyte, $d =$ 3 $\times$ 10$^{-10}$ m.
			{Legend: SPB$_{LH}$ (dashed red lines), MPB$_{LH}$ (dashed green lines), and
				MC$_{LH}$ (black open circles). DHLL (solid black line).} The rest of the physical parameters
			as in figure~\ref{fig1}.}
		\label{fig11}
	\end{center}
\end{figure}
\begin{figure}[!t]
	\begin{center}
		\includegraphics[width=0.5\textwidth]{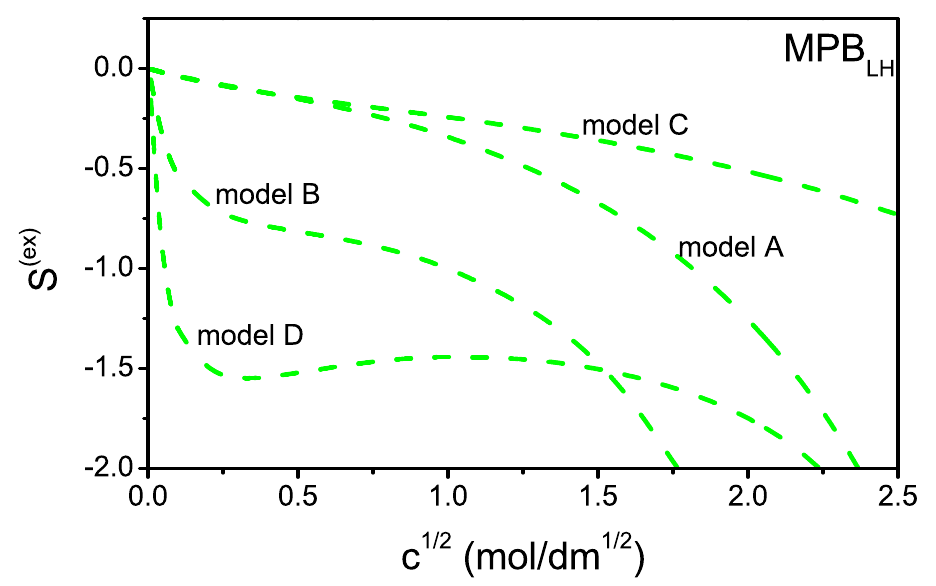}
		\caption{(Colour online) Excess entropy $S^{(\text{ex})}$ as a function of the
			square  root  of electrolyte concentration using the MPB$_{LH}$ for the models A, B, C, and D.}
		\label{fig12}
	\end{center}
\end{figure}
\begin{figure}[!t]
\begin{center}
	\includegraphics[width=0.5\textwidth]{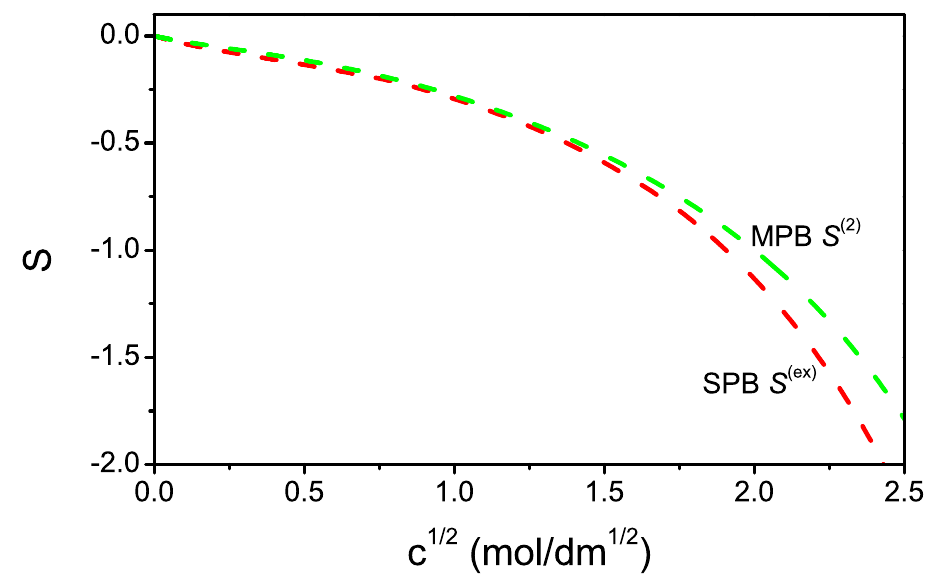}
	\caption{(Colour online) Comparison of  the  excess entropy, $S^{(\text{ex})}$,
		from the SPB theory with the $S^{(2)}$ component from the MPB theory as functions of the
		square  root  of electrolyte concentration. The physical parameters are from model A.
		Legend: SPB (red dashed line), MPB (dashed green line).}
	\label{fig13}
\end{center}
\end{figure}

    { A fundamental reason for the different dispositions of the TI and LH
results for the $S^{(\text{ex})}$ for the SPB, MPB, and MC in figures~\ref{fig2}--\ref{fig4} can be traced back to the
behaviour of the corresponding predicted  $g_{ij}(r)$ vis-\'{a}-vis those of the simulations. For
instance, in figure~\ref{fig1}  at a fairly high $c = 1.96$~mol/dm$^{3}$,  we saw that for 1:1 valency
case the SPB and the MPB profiles are quite reasonable relative to the simulations but
for 2:2 valency case there are discrepancies with the SPB, in particular, not being qualitative.
Such trends are likely to accentuate at still higher concentrations. This is manifested
in the $S^{(\text{ex})}$ results in these figures.  In figure~\ref{fig2} (1:1 case) the MC$_{TI}$ and MC$_{LH}$
are virtually indistinguishable.  The MPB$_{LH}$ underestimates the simulations at very high
concentrations (which may be attributed to the deviations in the MPB profiles) and remain
below the MPB$_{TI}$. Two other factors that are likely to impact more the 2:2 valency results
are:

\begin{enumerate}[(i)]
	\item the neglect of  the higher order terms beyond $S_{\text{ring}}$ in the LH expansion , and
 \item the fact that the $U$ and $\phi ^{(\text{ex})}$ terms in the TI expression [equation~(\ref{eq12})]
\end{enumerate}
involve only the contact values of the potential and distribution profiles, while $\gamma _{\pm}$ are 
calculated using the G\"{u}ntelberg charging process~\cite{Outh5,Outh5ol}, which gives the
expression somewhat of a local character. By contrast, in evaluating the $S^{(\text{ex})}$ from the LH
equations, the full range of the $g_{ij}(r)$ is necessary. This non-local nature is likely to
lead to more errors.

We note here that at low concentrations for 2:2 valency cases in figures~\ref{fig3} and \ref{fig4} the
MC$_{LH}$ $S^{(\text{ex})}$ $<$ MC$_{TI}$ $S^{(\text{ex})}$. The MPB$_{LH}$
and MPB$_{TI}$ follow the same trend, although differing numerically from the simulation data.
The SPB$_{LH}$ and SPB$_{TI}$, on the other hand, reveal a different trend with
SPB$_{LH}$ $S^{(\text{ex})}$ $>$ SPB$_{TI}$ $S^{(\text{ex})}$ over nearly the entire concentration
region probed. The physical reason for this may well be the effect
of the missing inter-ionic correlations in the classical mean field theory.

    We next compare our results with one set of results for $S^{(\text{ex})}$
for bulk electrolytes available in the literature. Hummer and Soumpasis \cite{Hummer} 
calculated the $S^{(\text{ex})}$ for NaCl solutions using the RPM in conjunction with HNC integral
equation theory, Widom's particle insertion MC technique, and an entropy expansion.
The physical parameters used by these authors were $d = 4.9 \times 10^{-10}$ m,
$\varepsilon _{r} = 78.356$, and $T = 298$~K. In figure~\ref{fig5}, we have plotted the
entropy results at these parameters from the SPB$_{TI}$ (solid red line), the
MPB$_{TI}$ (solid green line), the SPB$_{LH}$ (dashed red line), and the
MPB$_{LH}$ (dashed green line) again as functions of $\sqrt{c}$. The Hummer-Soumpasis
results from table~IV of reference \cite{Hummer} are shown as: $s^{(2)} + s^{(3)}$
(black stars), MC (Widom) (red filled circles), and HNC (blue squares). The SPB and
MPB curves compare reasonably well with the Hummer-Soumpasis data overall. The TI
results from SPB and MPB are almost quantitative with their MC values. The
discrepancy between the SPB$_{LH}$ or MPB$_{LH}$ and the $s^{(2)} + s^{(3)}$ results
is most likely due to the differences in the approximations involved in the entropy
expansion techniques.

    In the next part of the work (figure~\ref{fig6}--\ref{fig11}), we examine more closely how the
$g_{ij}(r)$'s for the four electrolyte models (A, B, C, and D) affect the $S^{(\text{ex})}$
as a function of  $\sqrt{c}$. Some of the $S^{(\text{ex})}$ results for the model systems
were seen in the figures~\ref{fig2}--\ref{fig5}, but are included here for completeness. Figure~\ref{fig6} shows
the $S^{(\text{ex})}$ and its constituent components $S^{(2)}$ and $S_{\text{ring}}$ for model A
in the low concentration range, where the DHLL is applicable. As expected, all the DHLL
curves are linear with respect to $\sqrt{c}$. It is not surprising that the SPB and the
MPB curves together with the simulation results are also almost linear trending to the
correct low concentration DHLL behaviour. Minor noise in the MC$_{LH}$ data points are
visible in the area of the lowest concentrations. This is presumably due to possible
imprecision in our calculations. At such low concentrations one needs to evaluate the
$g_{ij}(r)$'s around an ion out to very large distances, which can be challenging
numerically. The magnitude of the $S_{\text{ring}}$ component of the $S^{(\text{ex})}$ is much
smaller than the magnitude of the $S^{(2)}$ component. The lowest entropy values
are predicted by the DHLL theory followed by the MPB, MC and SPB. The results
for a wider range of concentrations are shown in figure~\ref{fig7} without DHLL. The remaining
curves are no longer linear with the theories still being in good agreement with the
simulations. Starting from a value of  $c\sim1$ mol/dm$^{3}$, a substantial decrease
in $S^{(\text{ex})}$ and in its components is visible, with the same ordering of the curves and
symbols to the entropy as in figure~\ref{fig6}.

    In model B, the ionic valency increases relative to that in model A.
The results for model B are shown in figures~\ref{fig8} and \ref{fig9}, at low and high
concentrations, respectively. Figure~\ref{fig8} reveals that even in the low
concentration regime, the curves, with the exception of the DHLL, are not
linear. The $S^{(2)}$ and $S^{(\text{ex})}$ curves are convex, while the $S_{\text{ring}}$
curves are initially concave becoming more linear at higher concentrations.
The theories are qualitative with the simulations, although the MPB results
tend to follow the MC data more closely than do the corresponding SPB results.
Note though that the scales in figure~\ref{fig8} are expanded and the same curves appear
closer together at the scales of figure~\ref{fig9} at low concentrations. Also in figure~\ref{fig9},
a sharp decrease in $S^{(2)}$, $S_{\text{ring}}$, and $S^{(\text{ex})}$ is visible at low
concentrations followed by an inflection point, after which the curves become
concave. The SPB curves pass through a maximum, the MC curves have a flat plateau,
and the MPB curves decrease monotonously. A similar pattern for $S^{(\text{ex})}$ for
divalent ions was observed by Laird and Haymet \cite{Laird-Haymet2}. This behavior is
somewhat similar to the dependence of the logarithm of the activity coefficient
on the square root of the concentration. At low concentrations, electrostatic
interactions dominate, which give a negative contribution to the activity coefficient,
while as the concentration increases, steric interactions start to become more relevant,
which make a positive contribution. It is not difficult to locate the boundary between
the regions of dominance of electrostatic and steric interactions. However, this is not
the case with entropy, where both types of interactions lead to a negative contribution.
This blurs the boundaries between these areas so that increasing the interactions as
in model B may be helpful here. Figure~\ref{fig9} clearly shows three areas: predominant electrostatic
interactions, transient interactions, and predominant steric interactions.

    The entropy curves for model C at low electrolyte concentrations are not
shown as they are very similar to those presented in figure~\ref{fig6} for model A.
There are similarities also between the sets of curves in figure~\ref{fig7} (model A)
and in figure~\ref{fig10} (model C). For instance, overall the curves follow the MC data.
The MPB is semi-quantitative or better, although the SPB $S_{\text{ring}}$ curve deviates
from the simulations at higher concentrations displaying a slight hump. The other
differences between the two models are: the magnitudes of the $S^{(2)}$, $S_{\text{ring}}$,
and $S^{(\text{ex})}$ are smaller in figure~\ref{fig10} than in figure~\ref{fig7}, and the $S_{\text{ring}}$, and
$S^{(\text{ex})}$ curves have a slight inflection. This is likely caused by stronger
electrostatic interactions as two ions can be at closer proximity to each other.

    There is a substantial change in the panorama in going from figure~\ref{fig9} (model B)
at $d = 4.25\times 10^{-10}$~m to figure~\ref{fig11} at $d = 3 \times 10^{-10}$~m.
The behaviour of the $S^{(\text{ex})}$ was explained earlier in relation to figure~\ref{fig4} and
the TI results. Here, the MC data for both $S^{(2)}$ and $S_{\text{ring}}$ also reveal a
sharp drop at low concentrations giving a minimum. The analogous SPB and MPB
curves also show a minimum, but a much shallower one. This behavior is again due
to the increase in electrostatic interactions stemming from a reduced ion size
compared to that in model B. The increase in concentration leads to the appearance
of a flat hump. Subsequently, all the curves show a downward trend.

    Figure~\ref{fig12} allows us to view the results from the point of view of steric interactions.
For example, models A and C differ in the size of the ions with model A having the larger ions.
This results in the $S^{(\text{ex})}$ curves of model A are running below those in C. In the case
of models B and D, the picture is less clear. It is obscured by strong inter-ion
electrostatic interactions. However, in the lower right corner one can still observe
that the curves for B begin to decline faster than those for D, which is clearly due
to greater steric interactions. The above analysis is based on the results obtained
from MPB. Analogous conclusions can be reached when analyzing the results of SPB and MC.

    Finally, it is interesting to note that the inclusion of fluctuation terms in the MPB
theory vis-\`{a}-vis the ring approximation in the LH equations can have similar
consequences. For example, figure~\ref{fig13} shows (for model A) the $S^{(\text{ex})}$ obtained from
the SPB theory and the $S^{(2)}$ component obtained from the MPB theory as functions of
$\sqrt{c}$. Note that the SPB curve includes the ring approximation rather then the
fluctuation terms, while the MPB curve includes the fluctuation terms rather then the
ring approximation. Both curves are very similar to each other.

\section{Conclusions}

    The achievement of this paper has been an attempt to explore the statistical
mechanical, potential approaches of SPB and MPB theories to calculate the entropy
of RPM electrolytes on the basis of (a) thermodynamic integration, and (b) the
LH equations \cite{Laird-Haymet1,Laird-Haymet2}.

    Although for charged fluids entropy is as important a thermodynamic
variable as the osmotic and activity coefficients are, calculation of
the entropy has received relatively less attention perhaps because of
the issues involved in such calculations. From a theoretical
perspective, an estimation of fluid entropy necessitates a knowledge
of the distribution functions to all orders, which is a difficult task.
Besides the thermodynamic integration, the LH entropy expansion
suggests a way forward by an approximation that partially accounts for
the contribution of higher order distributions. Thus, only the second
order distribution, that is, the pair-correlation or pair distribution
function, which is standard staple of any formal theory,
is needed explicitly. The earlier application of the LH procedure
\cite{Laird-Haymet2} involved an electrolyte model
with a soft repulsive core --- not quite the primitive model.

    The work here shows that overall for the RPM the SPB results
for $S^{(\text{ex})}$ are generally qualitative with that from the simulations
at lower concentrations, while the MPB results are semi-quantitative or better.
Such trends are consistent with the SPB and MPB characterizations of
thermodynamics, for instance, osmotic and activity coefficients of primitive
model electrolytes vis-\`{a}-vis the corresponding simulations results reported
earlier in the literature \cite{Quin1,Quin2}. Furthermore, the results
from these potential based theories at $d = 4.25 \times 10^{-10}$ m
for both 1:1 and 2:2 valencies show many similar characteristics as that
from the integral equation theory used by Laird and Haymet \cite{Laird-Haymet2},
although the electrolyte models are not identical.
The SPB and MPB formalisms together with thermodynamic integration or
LH equations are also seen to reproduce well some excess entropy
results of Hummer and Soumpasis \cite{Hummer} obtained using a MC Widom particle
insertion method, the HNC integral equation, and an entropy expansion.

     An aspect of this work has been comparing the LH expansion results for
the $S^{(\text{ex})}$, evaluated by the MC radial distribution functions, with that
found by TI via simulation. For 1:1 electrolytes, the agreement between the two
approaches is excellent, with MC$_{TI} \simeq MC_{LH}$. Deficiencies between the
two approaches arise, however, in the 2:2 case. For low concentrations with
$d = 4.25\times 10^{-10}$~m, the LH  expansion is fairly reasonable and is
qualitatively correct at the higher concentrations. With $d = 3 \times 10^{-10}$ m
the LH scheme is inappropriate at low concentrations and is only qualitatively
correct at higher concentrations.

     { The thermodynamic properties of 1:1 and 2:2 salts differ due to the
strong interionic forces for higher charges. Examples include the negative deviation
of the 2:2 electrolyte mean activity coefficient from the DHLL at high dilution,
compared to the positive deviation for the 1:1 case, with an analogous behaviour
for the osmotic coefficient \cite{Harned,Robinson,Guggenheim}.
Bjerrum \cite{Bjerrum-Dansk} introduced the idea of ion pairs which has provided the
inspiration for many a theoretical investigation involving ionic association \cite{Barthel,Holovko}.
Support for ion association has been given by integral equation theories, although the
unmodified HNC theory overestimates the pair and triple aggregates \cite{Rossky,Rogde}.
It thus seems that ionic association is playing a fairly significant role in the initial rapid
decrease in $S^{(\text{ex})}$ at high dilution seen in figures~\ref{fig3} and \ref{fig4},
the smaller radius leading to a deep minimum. The MPB accurately predicts the RPM 2:2
thermodynamic and structural simulation results at $d = 4.25 \times  10^{-10}$ m, with its
accuracy reducing as $d$ decreases, as well as those predictions related to the classical
approach at low concentrations \cite{Malatesta}. The SPB is less successful than the MPB in
predicting $S^{(\text{ex})}$. As the MPB includes fluctuation
terms, the interpretation of ionic association can be ascribed to the addition of fluctuation terms
into the mean field SPB theory. The contribution of higher terms in the LH expansion is difficult
to assess as these terms contain multi-particle correlation functions. As indicated in the
previous Section, for the 1:1 case at the treated parameters, the comparison of the
MC$_{TI}$ and MC$_{LH}$ data in figure~\ref{fig2} indicate that these higher terms make a small
or negligible contribution. Deviations occur between the simulation
results for 2:2 electrolytes. At $d = 4.25 \times  10^{-10}$ m, the MC$_{TI}$ and MC$_{LH}$
diverge at the higher concentrations, while with $d = 3 \times  10^{-10}$~m,
the difference persists throughout the concentration range and can be very pronounced
(figure~\ref{fig4}). The neglect of the higher order terms in the LH theory for 2:2 electrolytes
can be interpreted as leading to an increase in ion association, as ion size reduces,
at high dilution.}

     Within the present formulation of the SPB theory, it might be
feasible to incorporate a soft core potential. The $g_{ij}^{0}$ would need to be
replaced by the corresponding quantity for a soft potential. In case of the MPB,
however, the $L_{i}(u_{j})$ also depends on the ion-size. Another possible
extension of the present work can be envisaged through the use of a variable
dielectric constant. The dependence of the $\varepsilon _{r}$ on the solution
concentration and/or temperature is an experimental phenomenon \cite{Barthel}.
Some of us were involved earlier in a thermodynamic analysis of some alkali
halide solutions with a concentration or temperature dependent $\varepsilon _{r}$
using the SPB, MPB, and the MSA with encouraging results \cite{Quin2} .
We hope to build on the present study along these lines in future.

\section*{Acknowledgements}

    Professor Stanis{\l}aw Lamperski passed away before the paper was completed.
He conceived and carried out the majority of the computations, which were
kindly finished by Dr. Rafa{\l} Gorniak.
We acknowledge financial support to Professor Lamperski from Adam Mickiewicz University,
Pozna\'{n}, Poland.

\newpage
\ukrainianpart

\title{┼эЄЁюя│  хыхъЄЁюы│Єє т Ёрьърї юсьхцхэю┐ яЁшь│Єштэю┐ ьюфхы│ ч тшъюЁшёЄрээ ь я│фїюфє ёхЁхфэ№юую хыхъЄЁюёЄрЄшўэюую яюЄхэЎ│рыє}

\author{
	\framebox{╤. ╦рьяхЁёъ│}\refaddr{label1}, ╦. ┴. ┴єi э\refaddr{label2}, ╩. ┬. ╬єЄтрщЄ\refaddr{label3},
	╨. ├юЁэ ъ\refaddr{label1}  }

\addresses{
	\addr{label1} ╒│ь│ўэшщ Їръєы№ЄхЄ, ╙э│тхЁёшЄхЄ │ьхэ│ └фрьр ╠│Ўъхтшўр, тєы. ╧ючэрэё№ъюую єэ│тхЁёшЄхЄє, 8, 61-614 ╧ючэрэ№, ╧юы№∙р
	\addr{label2} ╦рсюЁрЄюЁ│  ЄхюЁхЄшўэю┐ Ї│чшъш, ╘│чшўэшщ Їръєы№ЄхЄ, ╙э│тхЁёшЄхЄ ╧єхЁЄю-╨│ъю
	\addr{label3} ┬ш∙р °ъюыр ьрЄхьрЄшъш Єр Ї│чшъш, єэ│тхЁёшЄхЄ ╪хЇЇ│ыфр, ╪хЇЇ│ыф S3 7RH, ┬хышъюсЁшЄрэ│ 
}

\makeukrtitle

\begin{abstract}
	\tolerance=3000%
	═рфыш°ъютр хэЄЁюя│  хыхъЄЁюы│Є│т т Ёрьърї юсьхцхэю┐ яЁшь│Єштэю┐ ьюфхы│ ЁючЁрїютє║Є№ё  ч тшъюЁшёЄрээ ь ьхЄюфє яюЄхэЎ│рыє чр фюяюьюую■ ёшьхЄЁшўэю┐ ЄхюЁ│┐ ╧єрёёюэр-┴юы№Ўьрэр Єр ьюфшЇ│ъютрэю┐ ЄхюЁ│┐ ╧єрёёюэр-┴юы№Ўьрэр. ╓│ ЄхюЁхЄшўэ│ я│фїюфш тшъюЁшёЄютє■Є№ё  є яю║фэрээ│ ч Ё│тэ ээ ь ёЄрЄшёЄшўэю┐ ЄхЁьюфшэрь│ъш,  ъх,  ъ яюърчрэю, ║ хът│трыхэЄэшь ЄхЁьюфшэрь│ўэюьє │эЄхуЁєтрээ■. ╤шёЄхьш хыхъЄЁюы│Є│т ч │юээшьш трыхэЄэюёЄ ьш 1:1 │ 2:2 Єр ч ф│рьхЄЁрьш $3 \times 10^{-10}$ ь │ 4,25 $\times 10^{-10}$ ь Ёючуы фр■Є№ё  т °шЁюъюьє ф│рярчюэ│ ъюэЎхэЄЁрЎ│щ. ╥юўэ│ Ёрф│ры№э│ ЇєэъЎ│┐ Ёючяюф│ыє фы  ьюфхы№эшї хыхъЄЁюы│Є│т, юЄЁшьрэ│ т Ёхчєы№ЄрЄ│ ьюфхы■трээ  ╠юэЄх-╩рЁыю т ърэюэ│ўэюьє рэёрьсы│, яюЁ│тэ■■Є№ё  ч т│фяют│фэшьш ЄхюЁхЄшўэшьш яхЁхфсрўхээ ьш. ╩Ё│ь Єюую, ЇєэъЎ│┐ Ёрф│ры№эюую Ёючяюф│ыє, юЄЁшьрэ│  ъ ЄхюЁхЄшўэшь ўшэюь Єръ │ ьхЄюфюь ьюфхы■трээ , тшъюЁшёЄютє■Є№ё  т Ё│тэ ээ ї Ёючъырфє хэЄЁюя│┐ ╦хщЁфр-╒рщьхЄр [J. Chem. Phys., 1994, \textbf{100}, 3775] фы  юЎ│эъш эрфыш°ъютю┐ хэЄЁюя│┐ Ёючўшэ│т. ╓│ Ё│тэ ээ  тЁрїютє■Є№ срурЄюўрёЄшэъют│ ЇєэъЎ│┐ Ёючяюф│ыє,  ъ│ ряЁюъёшьє■Є№ё  чр фюяюьюую■ ``ъ│ы№Ўхтюую'' фюфрэър. ╟рурыюь, ьюфшЇ│ъютрэр ЄхюЁ│  ╧єрёёюэр-┴юы№Ўьрэр фр║ Ёхчєы№ЄрЄш,  ъ│ с│ы№° єчуюфцє■Є№ё  ч фрэшьш ьюфхы■трээ , э│ц ч Ёхчєы№ЄрЄш ёшьхЄЁшўэю┐ ЄхюЁ│┐ ╧єрёёюэр-┴юы№Ўьрэр. ╬ЄЁшьрэ│ Ёхчєы№ЄрЄш яюърчє■Є№, ∙ю эрфыш°ъютр хэЄЁюя│  ║ т│фТ║ьэю■, р ┐┐ рсёюы■Єэх чэрўхээ  чЁюёЄр║ фы  хыхъЄЁюы│Є│т 1:1 ч│ чс│ы№°хээ ь ъюэЎхэЄЁрЎ│┐. ╤шьхЄЁшўэ│ чэрўхээ  ╧єрёёюэр-┴юы№Ўьрэр фх∙ю чртш∙хэ│, р ьюфшЇ│ъютрэ│ чэрўхээ  ╧єрёёюэр-┴юы№Ўьрэр фх∙ю чрэшцхэ│ є яюЁ│тэ ээ│ ч ьюфхы№эшьш. ╩Ёшт│ фы  хыхъЄЁюы│Є│т 1:1 тъы■ўэю ч ъЁштшьш, юЄЁшьрэшьш ч Ё│тэ э№ ╦хщЁфр-╒рщьхЄр, єчуюфцє■Є№ё  юфэр ч юфэю■, Єюф│  ъ ъЁшт│ фы  хыхъЄЁюы│Є│т 2:2 яЁш $4,25 \times 10^{-10}$ ь, юЄЁшьрэ│ т Ёрьърї ьюфшЇ│ъютрэю┐ ЄхюЁ│┐ ╧єрёёюэр-┴юы№Ўьрэр, ыш°х  ъ│ёэю єчуюфцє■Є№ё  ч фрэшьш ьюфхы■трээ  фю уєёЄшэ ~1 ьюы№/фь$^{3}$. ╩Ёшт│ фы  хыхъЄЁюы│Є│т 2:2 ьр■Є№ їрЁръЄхЁэшщ яхЁхушэ │ яырЄю. ╨хчєы№ЄрЄш, юЄЁшьрэ│ т ф│рярчюэ│ эшч№ъшї ъюэЎхэЄЁрЎ│щ~(<~0,01~ьюы№/фь$^{3}$), єчуюфцє■Є№ё  ч яхЁхфсрўхээ ьш чръюэє ─хср -├■ъъхы .
	\keywords хыхъЄЁюы│Єш, юсьхцхэр яЁшь│Єштэр ьюфхы№, хэЄЁюя│ , ЄхЁьюфшэрь│ўэх │эЄхуЁєтрээ , ьхЄюф ╠юэЄх-╩рЁыю, ёшьхЄЁшчютрэ│ ьюфшЇ│ъютрэ│ ЄхюЁ│┐ ╧єрёёюэр-┴юы№Ўьрэр
	
\end{abstract}
\end{document}